\documentclass[12pt,preprint]{aastex}

\usepackage{CJK}

\shorttitle{A magnetic reconnection origin for the soft X-ray excess in AGNs}
\shortauthors{Zhong \& Wang}

\begin{document}

\begin{CJK}{GB}{gbsn}

\title{A magnetic reconnection origin for the soft X-ray excess in AGN}

%
\author{Xiaogu Zhong\altaffilmark{1,2,3}, Jiancheng Wang\altaffilmark{1,2}}

\altaffiltext{1} {National Astronomical Observatories, Yunnan Observatory, Chinese Academy
of Sciences,  Kunming 650011, China}
\altaffiltext{2} {Key Laboratory for the Structure and Evolution of Celestial Objects,
Chinese Academy of Sciences,  Kunming 650011, China}
\altaffiltext{3} {Graduate School, Chinese Academy of Sciences, Beijing, P.R. China}

\email{guqian29@ynao.ac.cn}

\begin{abstract}
We present a new scenario to explain the soft X-ray excess in Active Galactic Nucleus. The magnetic reconnection could happen in a thin layer on the surface of accretion disk. Electrons are accelerated by shock wave and turbulence triggered by magnetic reconnection, then they take place inverse Compton scattering above accretion disk which contributes soft X-rays. Based on standard disk model, we estimate the magnetic field strength and the energy released by magnetic reconnection along accretion disk, and find that the luminosity caused by magnetic reconnection mainly emits in the inner disk which is dominated by radiation pressure. We then apply the model to fit the spectra of AGNs with strong soft X-ray excess.
\end{abstract}


\section{Introduction}

The soft X-ray excess is a major component of many type 1 active galactic nuclei (AGN), especially Narrow Line Seyfert 1 (NLS1) galaxies \citep[e.g.][]{bol96,lao97}. Below 1 $keV$ in the X-ray band, the X-ray data are above the low energy extrapolation of the best fitting 2-10 $keV$ power law by index about $\Gamma\sim2.0$ \citep{gon97}, this is so-called soft X-ray excess. Historically, the soft X-ray excess has often been fitted using a black body model, yielding best-fit temperatures in the range of 0.1-0.2 $keV$ \citep{cze03,gie04}. Standard disk model \citep{sha73} gives a maximum effective temperature of the accreting material of $kT\sim10(\frac{\dot{m}}{m_8})^{1/4}$eV, where $\dot{m}$ is the accretion rate in unit of Eddington accretion rate (e.g. $\dot{m}=\dot{M}/\dot{M_E}$) and $m_8=M/10^8M_\odot$. However, according to this model, only with high accretion rate and low-mass black holes, the peak of accretion disk blackbody spectrum is just at 0.1 $keV$. But the current observation shows that the black hole mass of source with soft X-ray excess exceeds $10^6M_\odot$ \citep{cru06}. Because different NLS1 have different mass and accretion rate, the blackbody spectrum of accretion disk is difficult to maintain a stable temperature within the range of 0.1-0.2 $keV$. Standard disc model is difficult to explain these properties \citep{tur89,don12}.

From the analysis of the soft X-ray data, it is found that strong emission and absorption features appear in the soft X-ray spectra of most AGN. As discussed by \citet{gie04}, the imprint of spectral features such as the strong jump in opacity at $\sim0.7$keV from $\texttt{O}_{VII}$, $\texttt{O}_{VIII}$ and Fe M-Shell can lead to an apparent soft excess below this energy. However, the problem is that some parts of the soft excess is often smooth, and certain predicted features are not evident. \citet{sch07,sch08} computed the X-ray spectra of outflow gas with density and velocity, demonstrating that high outflow velocity would be required if absorption in outflows solely explain soft excesses observed in AGN spectra. The high velocity with 0.9 c even exceeds the relativistic wind components detected through energy shifted absorption lines, and \citet{sch09} have found that the models of line-driven accretion-disk winds do not attain sufficiently high velocities.

The origin of the soft X-ray excess in AGNs is as yet unknown, we try to explore new radiation mechanism through considering the roles of the magnetic field in accretion disks. We present a new scenario follow as: the magnetic reconnection could happen in a thin layer on the surface of accretion disk like solar active region \citep{kur84}. Electrons are accelerated by shock wave and turbulence triggered by magnetic reconnection \citep{kir94}, then these electrons emit the X-rays through Compton scattering the photons from the accretion disk. In Section 2, we estimate the magnetic field and the energy released by magnetic reconnection on accretion disks to show whether the released magnetic energy  can provide the radiation of soft X-ray excess. In Section 3, we depict the radiative processes and obtain the Compton spectrum. In Section 4, we apply new model to AGNs with strong soft X-ray excess. Finally, Section 5 summarizes our main results and presents our conclusions.

\section{Magnetic Field And Energy Release}

We assume that the magnetic reconnection emerges frequently in a thin-layer on the surface of standard accretion disk. In the past works, the calculation of the magnetic field $B$ adopts the equipartition of gas energy and magnetic energy in the disk given by \citet{mil00}, $\xi\equiv n_{disk}kT_{disk}/(B^2/8\pi)$, but this result is not a useful guide because they ignore radiation and assume isothermal equation of state, and don't give how the magnetic field scales in the radiation dominated regime. In fact, it is unclear why the magnetic energy density should scale as a constant fixed ratio of the midplane gas pressure. The best numerical simulations have been done by \citet{hbk09,hkb09} and \citet{bla11}. Their simulations show that the vertically averaged rms of the magnetic field scales as the square root of the total (gas + radiation) pressure, e.g.
\begin{eqnarray}
B=\sqrt{8\pi \xi^{-1}(nkT+\frac{aT^4}{3})},
\end{eqnarray}
where $a=7.564\times10^{-15} erg\cdot cm^{-3}K^{-4} $ . Fig. 1 show the distribution of the magnetic field against the radius of accretion disc in unit of $R_S$ with the parameters of $\alpha=0.1$, $\xi=1$ and $10^8M_\odot$. It is shown that the magnetic field rapidly declines  along the radius of the accretion disk inside about 20$R_S$, but is relatively flat outside 20$R_S$, in which the different curves are calculated at the accretion rates $\dot{m}$ of 0.01, 0.1 and 1 respectively ($\dot{m}$ is in the unit of Eddington accretion rate). The magnetic field in the inner disk is stronger than that in the outer disk, implying that the magnetic energy is mainly deposited in the inner disk. We calculate the energy released by fast magnetic reconnection on the accretion disk, in which the magnetic dissipation scales as $B^2V_A$, and $V_A$ is the Alfven speed. But it has an upper limit near the disk surface, which is limited by the Poynting flux (see Fig.2) as \citep{hir06}:
\begin{eqnarray}
F_p=fc_s\frac{B^2}{8\pi},
\end{eqnarray}
where the factor $f$ represents the departure from hydrostatic balance which permits the upward motion, and $c_s$ is the sound velocity. From Fig.2, we know that the Poynting flux rapidly decreases from the innermost steady radius to 20$R_S$. Because the gas pressure is far less than the radiation pressure inside 20$R_S$, the Poynting flux is approximately given by:
\begin{eqnarray}
F_p\approx4.98\times10^{18}(\frac{4}{3}+\frac{2}{\xi})^{\frac{1}{2}}
\alpha^{-1}m_8^{-1}\dot{m}
{(\frac{R}{R_S})}^{-3}f\xi^{-1} erg\cdot cm^{-2}s^{-1},
\end{eqnarray}
where $R_S$ is the Schwarzschild radius. Assuming almost all energy released by magnetic reconnection be converted to soft X-ray radiation energy, we can estimate the maximum of soft X-ray luminosity by integrating the Poynting flux from the inner radius to 20$R_S$ as:
\begin{eqnarray}
L_{Soft}\approx2.81\times10^{46}(\frac{4}{3}+\frac{2}{\xi})^{\frac{1}{2}}
\alpha^{-1}m_8^{2}\dot{m}f\xi^{-1} erg\cdot s^{-1}.
\end{eqnarray}



\section{The Radiative Processes}
\subsection{The electron spectrum}
Considering that electrons are accelerated by shock wave and turbulence in magnetic reconnection, we can obtain the electron energy spectrum as \citep{dro86}:
 \begin{eqnarray}
F(E)=AE^{-q(E)},
\end{eqnarray}
where the electron spectrum index is
\begin{eqnarray}
q(E)=-\frac{\eta+a+1}{2}+\frac{3+a-\eta}{2}[1+
\frac{4\lambda{\sqrt{2m_e}}^{\eta+b}}{{(3+a-\eta)}^2}E^{(\eta+b)/2}]^{1/2},
\end{eqnarray}
and the parameters $\eta$, $a$, $\lambda$ and $b$ are determined by the microcosmic structure of the shock wave and turbulence expressed as follows \citep{bog85,dro86}:
\begin{eqnarray}
T(p)=T_0p^{-b},\nonumber \\
K(p)=\delta \upsilon (p)p^{2-q}=\kappa p^{\eta},\nonumber \\
D(p)=\alpha_2\frac{V_A^{2}}{K}p^{2},\nonumber \\
\dot{p_G}\approx \alpha_1\frac{V_S^{2}}{K}p^{2}, \nonumber
\end{eqnarray}
where $a=\alpha_1{V_S}^{2}/\alpha_2{V_A}^{2}$, $\lambda=\kappa/\alpha_2{V_A}^{2}T_0$. When the electrons are not accelerated to high energy for fast magnetic reconnection, the electron spectrum will become a power-law with constant index \citep{dro86}, given by
\begin{equation}
F(E)=AE^{-q}, E_1\leq E\leq E_2;
\end{equation}
$E_1$ and $E_2$ are the cutoffs in the spectrum. When $E_2\gg E_1$ and $q>1$, the factor $A$ is given by $A=N_e(q-1)E_1^{q-1}$, in which $N_e$ is the number density of electrons.
The kinetic energy $E$ of non-relativistic electron is expressed as $E=(\gamma-1)m_ec^2$, in which $\gamma=1/\sqrt{1-\beta^2}$ and $\beta c$ are the Lorentz factor and velocity of the electron. When $E\ll m_ec^2$, $\beta=\sqrt{2E/m_ec^2}$.

\subsection{Comparison of electron energy losses through Bremsstrahlung and Compton radiation}
The electron in magnetic field will take place bremsstrahlung, inverse Compton scattering and synchrotron emission.
Because the electron is non-relativistic, we can ignore synchrotron radiation.
Next, we will compare the electron energy losses through bremsstrahlung and Compton radiation.

The energy loss rate of the electron with the energy $E$ in Compton radiation is given by \citep{blu70}
\begin{eqnarray}
\frac{dE}{dt}=\sigma_Tc\gamma^2\int(1-\beta cos\theta)^2\epsilon dn=\sigma_Tc\gamma^2(1+\frac{1}{3}\beta^2)U_{ph},
\end{eqnarray}
where $\sigma_T$ is the Thompson cross section, $U_{ph}$ is the energy density of soft photons. The total energy loss rate of the non-thermal electrons is given by
\begin{eqnarray}
P_c&=&\sigma_Tc U_{ph}\int^{E_2}_{E_1}A\gamma^2(1+\frac{1}{3}\beta^2)E^{-q}dE,\nonumber \\
&=&\sigma_Tc U_{ph}N(q-1)(\frac{E_1}{m_ec^2})^{q-1}\int^{\beta_2}_{\beta_1}\gamma^5\beta(1+\frac{1}{3}\beta^2)(\gamma-1)^{-q}d\beta.
\end{eqnarray}

The energy loss rate of the electron with the energy $E$ in Bremsstrahlung radiation is given by \citep{gou80}
\begin{equation}
\frac{dE}{dt}=\frac{2}{\pi}\sigma_T n_Z Z^2 \alpha_{f}m_ec^3\beta(1+\frac{5}{12}\beta^2),
\end{equation}
$\alpha_f=\frac{1}{137}$ is the fine structure constant, $n_Z$ is the number density of the ion. The total energy loss rate is
\begin{eqnarray}
P_b&=&\sum_Zn_Z Z^2\frac{2}{\pi}\sigma_T \alpha_{f}m_ec^3A\int^{E_2}_{E_1}\beta(1+\frac{5}{12}\beta^2)E^{-q}dE,\nonumber\\
&=&\sum_Zn_Z Z^2\frac{2}{\pi}\sigma_T  \alpha_{f}m_ec^3N(q-1)(\frac{E_1}{m_ec^2})^{q-1}\int^{\beta_2}_{\beta_1}\gamma^3\beta^2(1+\frac{5}{12}\beta^2)(\gamma-1)^{-q}d\beta.
\end{eqnarray}
Then we have
\begin{eqnarray}
\frac{P_c}{P_b}=\frac{\pi U_{ph}}{2\alpha_f n_p m_ec^2}\frac{H_1(\beta_1,\beta_2)}{H_2(\beta_1,\beta_2)},
\end{eqnarray}
where $n_p=\sum n_Z Z^2$,
\begin{equation}
H_1(\beta_1,\beta_2)=\int^{\beta_2}_{\beta_1}\gamma^5\beta(1+\frac{1}{3}\beta^2)(\gamma-1)^{-q}d\beta,
\end{equation}
and
\begin{eqnarray}
H_2(\beta_1,\beta_2)=\int^{\beta_2}_{\beta_1}\gamma^3\beta^2(1+\frac{5}{12}\beta^2)(\gamma-1)^{-q}d\beta.
\end{eqnarray}
In the accretion disk, $U_{ph}=\sigma T_{eff}^4/c$, $T_{eff}$ is the effective temperature of disk, $\sigma$ is the radiation constant, and $n_p$ takes the number density of the proton, we obtain
\begin{eqnarray}
\frac{P_c}{P_b}=\frac{\pi \sigma T_{eff}^4}{2\alpha_f n_p m_ec^3}\frac{H_1(\beta_1,\beta_2)}{H_2(\beta_1,\beta_2)}=1.78\times 10^7 \alpha \dot{m}^3 r^{-\frac{9}{2}}(1-r^{-\frac{1}{2}})^2 \frac{H_1(\beta_1,\beta_2)}{H_2(\beta_1,\beta_2)},
\end{eqnarray}
where $r=R/3R_s$, $\dot{m}=\dot{M}/\dot{M}_{Edd}$, and $\dot{M}_{Edd}$ is the Eddington accretion rate. Because $H_1(\beta_1,\beta_2)>H_2(\beta_1,\beta_2)$, $P_c\gg P_b$ in the inner region of accretion disk for larger accretion rate, indicating
that the Compton radiation of electrons dominates the bremsstrahlung radiation.

\subsection{Compton radiation}
Because the magnetic field decreases along the radius of accretion disk, the parameters of accelerated electrons will change with the radius. Assuming that $E_1$, $E_2$ and $q$ are independent of the magnetic field, we can estimate the number density of electrons along the radius.
Assuming that the energy loss rate of electrons equals the magnetic dissipation, e.g., $P_c=\frac{B^2 V_A}{8\pi h_r}$, in which $h_r$ is the characteristic scale of reconnection, we obtain
\begin{equation}
N(R)(q-1)(\frac{E_1}{m_ec^2})^{q-1}=\frac{B^2 V_A}{8\pi h_r \sigma_Tc U_{ph}}H_1^{-1}(\beta_1,\beta_2).
\end{equation}

When an electron is not accelerated to a high energy, the Compton spectrum in the Thomson limit given by \citet{blu70} for a relativistic electron should be changed as
\begin{eqnarray}
\frac{dN_{\gamma,\epsilon}}{dtd\epsilon_1}=\frac{\pi r_0^2 c}{2\gamma^4 \beta^4}\frac{n(\epsilon)d\epsilon}{\epsilon^2}
\left\{2\epsilon_1 ln\frac{\epsilon_1}{(1+\beta)^2 \gamma^2 \epsilon}+\frac{2\beta}{1+\beta}\epsilon_1+(1+\beta)(1+\beta^2)\gamma^2\epsilon-\frac{\epsilon_1^2}{(1+\beta)\gamma^2\epsilon}\right\},
\end{eqnarray}
where $n(\epsilon)d\epsilon$ represents the differential photon density, $\epsilon_1$ expresses the energy of the scattered photon, and the electron has the energy $E=(\gamma-1)m_e c^2$ and velocity $\beta c$.
When $\beta\rightarrow 1$, the above formula returns to the Compton spectrum produced by a high energy electron in the Thomson limit \citep{blu70}. It is noted that the above distribution function equals zero at $\beta_*=(\epsilon_1/\epsilon-1)(\epsilon_1/\epsilon+1)^{-1}$.
Since the differential number of electrons are $dN=AE^{-q}dE$, the total Compton spectrum is given by
\begin{eqnarray}
\frac{dN_{\epsilon}}{dt d\epsilon_1}=A\int E^{-q}dE \frac{dN_{\gamma,\epsilon}}{dtd\epsilon_1}=N_e(R)(q-1)(\frac{E_1}{m_ec^2})^{q-1}\int_{\beta_{min}}^{\beta_{max}}
(\gamma-1)^{-q}\gamma^3\beta d\beta\frac{dN_{\gamma,\epsilon}}{dtd\epsilon_1},
\end{eqnarray}
where $d\gamma=\gamma^3\beta d\beta$, $\beta_{min}=max[\beta_*,\beta_1]$, $\beta_{max}=\beta_2$, $\beta_1$ and $\beta_2$ correspond to the energy cutoffs $E_1$ and $E_2$ in the electron spectrum. For the soft photons in accretion disk, the differential photon density $n(\epsilon,R)d\epsilon$ in the radius $R$ is given by
\begin{eqnarray}
n(\epsilon,R)d\epsilon=\frac{8\pi\epsilon^2}{h^3c^3}\left(e^{\frac{\epsilon}{kT(R)}}-1\right)^{-1}d\epsilon,
\end{eqnarray}
where $T(R)$ is the temperature of disk surface in the radius $R$. We then obtain the Compton spectrum along the radius
\begin{eqnarray}
\frac{dN}{dtd\epsilon_1}(R)&=&N_e(R)(q-1)(\frac{E_1}{m_ec^2})^{q-1}\frac{\pi r_0^2 c}{2}\int_{\epsilon_0}^{\epsilon_1}\frac{n(\epsilon,R)}{\epsilon}H\left(\frac{\epsilon_1}{\epsilon},q\right)d\epsilon \nonumber\\
&=&\frac{3}{16}\frac{B^2 V_A}{8\pi h_r U_{ph}}H_1^{-1}(\beta_1,\beta_2)\int_{\epsilon_0}^{\epsilon_1}\frac{n(\epsilon,R)}{\epsilon}H\left(\frac{\epsilon_1}{\epsilon},q\right)d\epsilon,
\end{eqnarray}
where
\begin{eqnarray}
H\left(\frac{\epsilon_1}{\epsilon},q\right)=\int_{\beta_{min}}^{\beta_2}
(\gamma-1)^{-q}\gamma^{-1}\beta^{-3} d\beta \zeta\left(\frac{\epsilon_1}{\epsilon},\beta\right),
\end{eqnarray}
\begin{eqnarray}
\zeta\left(\frac{\epsilon_1}{\epsilon},\beta\right)=\frac{\epsilon_1}{\epsilon}
\left\{2ln\frac{\epsilon_1}{\epsilon}\frac{1}{(1+\beta)^2 \gamma^2}+\frac{2\beta}{1+\beta}+\frac{\epsilon}{\epsilon_1}(1+\beta)(1+\beta^2)\gamma^2-\frac{\epsilon_1}{\epsilon}\frac{1}{(1+\beta)\gamma^2}\right\},
\end{eqnarray}
and $\epsilon_0$ is given by $(\epsilon_1/\epsilon_0-1)(\epsilon_1/\epsilon_0+1)^{-1}=\beta_2$, e.g., $\epsilon_0=\epsilon_1(1-\beta_2)/(1+\beta_2)$.
In fact, $H\left(\frac{\epsilon_1}{\epsilon},q\right)$ can be simplified as
\begin{eqnarray}
H\left(\frac{\epsilon_1}{\epsilon},q\right)&=&\int_{\beta_1}^{\beta_2}
(\gamma-1)^{-q}\gamma^{-1}\beta^{-3} d\beta \zeta\left(\frac{\epsilon_1}{\epsilon},\beta\right), 1 \leq \frac{\epsilon_1}{\epsilon} <\frac{1+\beta_1}{1-\beta_1},\nonumber \\
H\left(\frac{\epsilon_1}{\epsilon},q\right)&=&\int_{\beta_*}^{\beta_2}
(\gamma-1)^{-q}\gamma^{-1}\beta^{-3} d\beta \zeta\left(\frac{\epsilon_1}{\epsilon},\beta\right), \frac{1+\beta_1}{1-\beta_1}\leq \frac{\epsilon_1}{\epsilon} \leq \frac{1+\beta_2}{1-\beta_2},\nonumber \\
H\left(\frac{\epsilon_1}{\epsilon},q\right)&=&0, \frac{\epsilon_1}{\epsilon} > \frac{1+\beta_2}{1-\beta_2}.
\end{eqnarray}

The total Compton radiation spectrum is given by
\begin{eqnarray}
L({\epsilon_1})&=& \int_{R_{in}}^{R_{out}} 2\pi R h_r \epsilon_1^2 \frac{dN}{dtd\epsilon_1}(R) dR \nonumber\\
&=&\frac{3}{64}H_1^{-1}(\beta_1,\beta_2)\epsilon_1^2 \int_{R_{in}}^{R_{out}} \frac{B^2 V_A}{ U_{ph}}\int_{\epsilon_0}^{\epsilon_1}\frac{n(\epsilon,R)}{\epsilon}H\left(\frac{\epsilon_1}{\epsilon},q\right)d\epsilon R dR,
\end{eqnarray}
where we take $R_{in}=3R_S$ and $R_{out}=20R_S$ for the radiation-dominated region of accretion disk, and $B^2 V_A/U_{ph}$ is given by
\begin{eqnarray}
\frac{B^2 V_A}{ U_{ph}}&=&\frac{8\pi c}{3 \sigma} \xi^{-\frac{3}{2}} \left(\frac{2a^3}{3m_p n_p}\right)^{\frac{1}{2}} \frac{T^6}{T_{eff}^4} \nonumber\\
&=&3.6\times 10^9 \alpha^{-1} \xi^{-\frac{3}{2}} (1-r^{-\frac{1}{2}})cm s^{-1},
\end{eqnarray}
where we use the structure of accretion disk as
\begin{eqnarray}
T(R)&=&2.3\times 10^5 (\alpha m_8)^{-\frac{1}{4}} r^{-\frac{3}{8}}K, \nonumber\\
n_p(R)&=&4.3\times 10^{9} (\alpha m_8)^{-1} \dot{m}^{-2} r^{\frac{3}{2}}(1-r^{-\frac{1}{2}})^{-2}cm^{-3}, \nonumber\\
T_{eff}(R)&=&6.2\times 10^5 (\dot{m}^{-1} m_8)^{-\frac{1}{4}} r^{-\frac{3}{4}}K, \nonumber
\end{eqnarray}
and $m_8=M/10^8 M_{\odot}$.

Finally we obtain
\begin{equation}
L(\bar{\epsilon})=6.9\times 10^{48} \alpha^{-1} \xi^{-\frac{3}{2}} m_8^2 \bar{L}(\bar{\epsilon},q,\beta_1,\beta_2) erg s^{-1},
\end{equation}
where $\bar{\epsilon}=\epsilon_1/keV$, and
\begin{equation}
\bar{L}(\bar{\epsilon},q,\beta_1,\beta_2,kT_0)=\bar{\epsilon}^2 H_1^{-1}(\beta_1,\beta_2)\int_{1}^{r_{out}}(1-r^{-\frac{1}{2}})rdr
\int_{\bar{\epsilon}_0}^{\bar{\epsilon}} \epsilon \left(e^{\frac{\epsilon}{kT_0}r^\frac{3}{8}}-1\right)^{-1}
H\left(\frac{\bar{\epsilon}}{\epsilon},q\right) d\epsilon,
\end{equation}
where $kT_0=2.0\times10^{-2}(\alpha m_8)^{-\frac{1}{4}}$, which means the temperature of inner stable radius of accretion disk,
and $r_{out}=20/3$,
The curves of $\bar{L}(\bar{\epsilon},q,\beta_1,\beta_2,kT_0)$ with $\bar{\epsilon}$ for the parameters of $q,\beta_1,\beta_2$ and $kT_0$
are shown in Figure 3-5.

\section{Model Application}

In this section we apply the model to fit the X-ray spectra (EPIC-pn of XMM-Newton data) of AGNs (PG quasars and NLS1s) with high mass accretion rates given by \citet{cru06} (see Table.1), in which their soft X-ray radiation include thermal radiation and Inverse Compton scattering. The Compton spectrum caused by magnetic reconnection is added in $XSPEC$ v12.7.1 \citep{arn96} through $initpackage$ and $lmod$ codes. Due to these samples have different black hole masses and accretion rates, they can get different temperature of the blackbody spectrum of accretion disk. According to the Standard disc model, for high accretion rate and low-mass black holes, the peak of accretion disc blackbody spectrum can arrive at X-ray band and we need to add the blackbody model in spectral fitting. Because the radiation of accretion disk is multi-temperature blackbody spectrum, we use the $diskbb$ model of the XSPEC to fit thermal spectrum.
The powerlaw with spectral index $\Gamma$ is used to fit the hard X-ray component because the hard X-ray radiation is considered to be from inverse Compton scattering in the accretion disc corona~\citep{liu02}. For some sources with low black hole mass and high accretion rate, the thermal radiation of accretion disk partly contributes the soft X-rays , we use the mixing model $absorption(diskbb+compton+powerlaw)$ to fit soft excess and use the $newpar$ code to link $kT_0$ and the temperature of inner stable radius of $diskbb$. For other sources with high black hole mass and low accretion rate, the thermal radiation can be ignored due to low blackbody temperature to radiate soft X-rays, we use the mixing model $absorption(compton+powerlaw)$ to fit soft excess. In the Compton model, there are four parameters to determine the spectrum, such as $kT_0$, $q$, $\beta_1$ and $\beta_2$. The fitting parameters are listed in Table 2, and the fitting spectra are shown in Fig.6-7. Then, we can estimate $\xi$ by equaling the observed soft X-ray luminosity to the Compton luminosity obtained by integrating the equation (26) from 0.1$keV$ to 5$keV$. Combining the equation (4) that gives the maximum of soft X-ray luminosity, we can estimate the maximum magnetic energy conversion rate, in which we take $f=0.1$ because the mean growing speed of magnetic structures is $0.1c_s$ \citep{hir06}. The results are listed in Table 3.

\section[]{Discussion and Conclusions}

The soft X-ray excess is a major component of the X-ray spectra in many AGNs, and is usually well fitted with a blackbody which has a roughly constant temperature of 0.1-0.2 keV over several decades of AGN mass. If the soft X-rays is thermal, its temperature is too high to be explained by the standard accretion disc model. Although the high temperature can occur in a slim accretion disc \citep{min00}, its accretion rate is super-Eddington \citep{tan05} and extreme ultraviolet photons are comptonized \citep{por04}. The strong emission and absorption features are found in the soft X-ray spectra of most AGNs. There is a strong jump in opacity at 0.7 keV from partially ionized material, where OVII/OVIII and Fe M shell unresolved transition array are combined to produce an apparent soft excess which is not intrinsic in two different geometries, either by reflection from optically thick material out of the line of sight, or through absorption by optically thin material in the line of sight. However, both geometries would show characteristic and sharp atomic features which can be smeared by strong relativistic effects in the reflection model, but the parameters required are quite extreme, in which the disc has to extend to the last stable orbit in high spin black hole, and the reflection is highly concentrated in the innermost regions \citep{cru06}. Reflection
off the surface of the accretion disc is proposed by \citet{fab02} to explain the soft excess, in which the primary X-rays go through substantial reprocessing in the upper layers of the disc, and emerge as a smooth spectrum due to relativistic blurring. Some sources are explained by disc reflection, such as, Mrk 335 \citep{lar08}, Mrk 478 \citep{zog08}, Fairall 9 \citep{schm09,emm11} and Ark 120 \citep{nar11}. A self-consistent, dual-reflector model is needed to explain the soft excess of Ark 120 at the different epochs \citep{nar12}.
Recently the intensity-correlated spectral analysis is developed to constrain the origin of the soft X-ray excess in several AGNs \citep{nod11,nod13}, in which the Multi-Zone Comptonization is required to interpret a stable soft excess component that is independent of the dominant power-law emission.  \citet{meh11,pat13} find that the soft X-ray excess emission varies in association with the UV emission and is independent of the X-ray power-law component based on the UV and X-ray observations of the Mrk 509. A long-look Suzaku observation of the Mrk 509 also support that the soft excess is independent of the X-ray continuum \citep{riv12}.  It is indicated that the soft X-ray excess could be produced by the Comptonisation of the thermal optical-UV photons from the accretion disc by a warm corona \citep{meh11} or with the disc photosphere itself \citep{don12}. \citet{don12} also find that the Compton upscattered soft X-ray excess decreases in importance with increasing $L/L_{Edd}$, in which the strongest soft excesses are associated with low mass accretion rate AGN rather than being tied to some change in disc structure around Eddington.

Based on the standard disk model, we present a new scenario to explain the soft X-ray excess. The magnetic reconnection is assumed to happen in a thin layer on the surface of accretion disk, its releasing energy accelerates electrons through shock wave and turbulence triggered by itself. Theses electrons then take place Inverse Compton scattering above accretion disk which contributes soft X-ray excess. We present that the magnetic field declines continuously with the radius of the accretion disk and decreases rapidly in the inside of disk, but is relatively flat in the outside of disk. We then calculate the upper limit of energy released by the Poynting flux on the accretion disk. The distribution of the Poynting flux implies that the soft X-ray excess should emit from the inner region of disk. In the inner region of accretion disk, the gas pressure can be ignored, we estimate the maximum luminosity by integrating the Poynting flux from the innermost steady radius to $20R_S$. In Section.3, we compare the electron energy losses through Bremsstrahlung and Compton radiation. The results show that the electron energy losses through Bremsstrahlung is far less than through Compton radiation. This means that the Compton radiation of electrons dominates the bremsstrahlung radiation. Then, we calculate the Compton spectrum of non-relativistic and non-thermal electrons, and obtain that the Compton spectrum is determined by $kT_0$, $q$, $\beta_1$, $\beta_2$ and is independent of accretion rate as found by \citet{win09}. From the equation (27), we know that the Compton spectrum is mainly determined by $kT_0$. According to the standard disc model, $T_0\propto M^{-1/4}$ and the masses of black hole in AGN are range of $10^6M_{\odot}$ to $10^9M_{\odot}$, the Compton spectrum weakly depends on mass. Fig.3 shows that the Compton spectrum with different $kT_0$ maintain a narrow range of 0.1-0.5 $keV$ corresponding to the characteristic temperature of the soft X-ray excess. Therefore, the Compton spectrum can well fit the soft X-ray excess as shown in Fig.6-7, in which Fig.7 includes the component with multi-temperature blackbody spectrum. It is shown that the soft photons from accretion disk are scattered to higher energy by non-thermal electrons generated by magnetic reconnection and contribute the soft excess. We also give the Compton spectrum depending on $q$, $\beta_1$ and $\beta_2$, shown in Fig.4-5. The spectrum keeps similar shape and shifts to low energy with $q$ increasing, while the spectrum also shifts in the energy with $\beta_1$ and $\beta_2$. Finally, we estimate the $\xi$ representing magnetic strength and the magnetic energy conversion rate listed in Table 3, and find that the magnetic energy conversion rates are low for many sources.

\section*{Acknowledgments}

We are grateful to the referee for useful comments improving the paper. We acknowledge the financial supports from the National Basic Research Program of China (973 Program 2009CB824800), the National Natural Science Foundation of China 11133006, 11163006, 11173054, and the Policy Research Program of Chinese Academy of Sciences (KJCX2-YW-T24).



\clearpage

\begin{figure}
\centerline{
    \includegraphics[scale=0.7,angle=0]{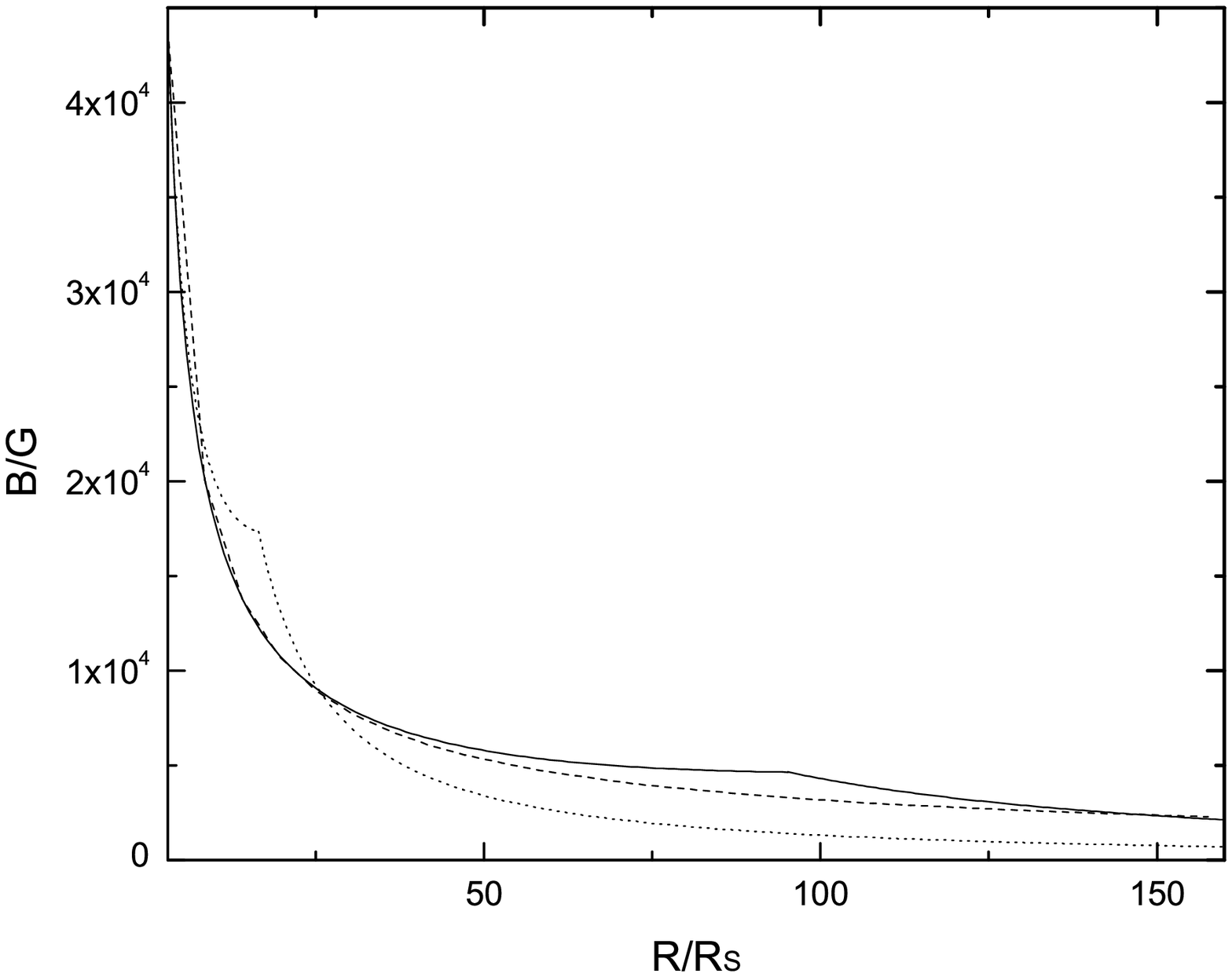}}
  \caption{The magnetic field against the radius of accretion disc in unit of $R_S$ with $\alpha=0.1$, $\xi=1$ and $10^8M_\odot$. The dash, solid and dot lines correspond to $\dot{m}$ of 1, 0.1 and 0.01 respectively.}
\end{figure}

\begin{figure}
\centerline{
    \includegraphics[scale=0.7,angle=0]{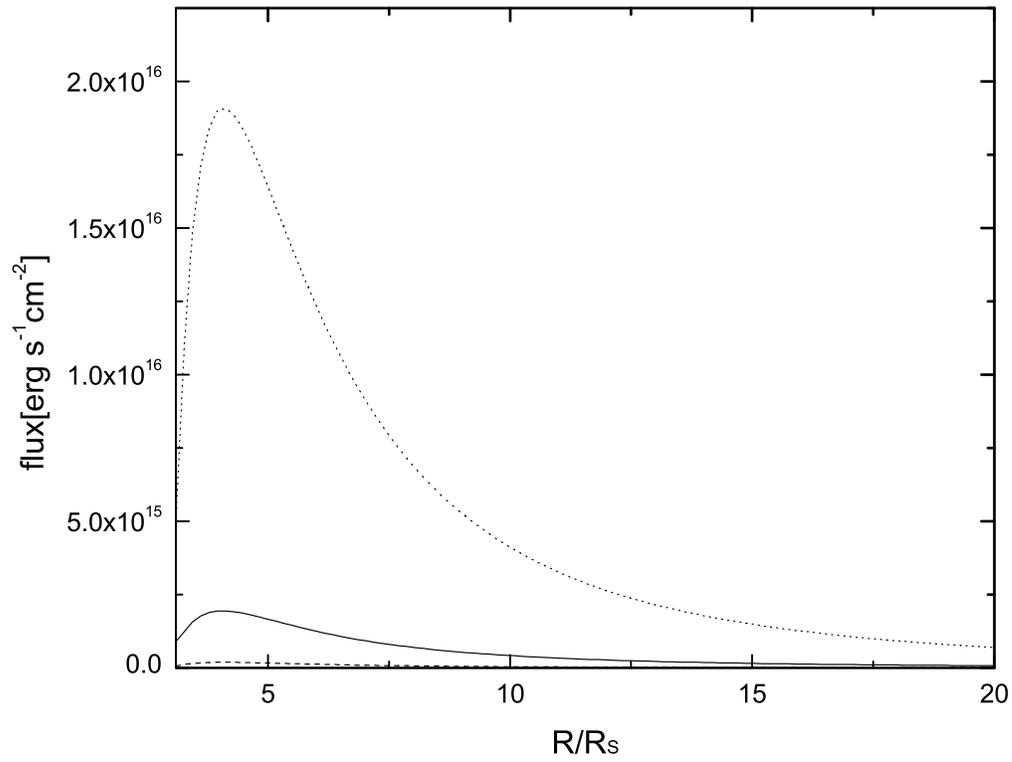}}
  \caption{The solid and dot lines represent the Poynting flux with $\alpha=0.1$, $\xi=1$, $10^8M_\odot$, $\dot{m}=0.1$, and  $\alpha=0.1$, $\xi=1$, $10^8M_\odot$, $\dot{m}=1$, respectively. The dash line represent the Poynting flux with $\alpha=0.1$, $\xi=1$, $10^9M_\odot$, $\dot{m}=0.1$}
\end{figure}

\begin{figure}
\centerline{
    \includegraphics[scale=0.5,angle=-90]{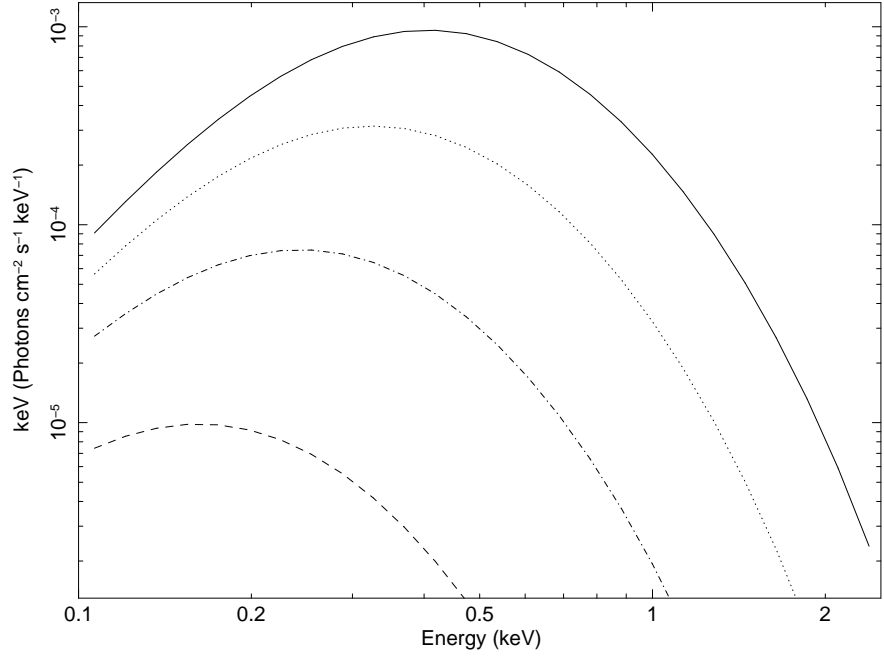}}
  \caption{The spectral shapes with different $kT_0$, in which the dash, dash dot, dotted and solid lines correspond to $kT_0=0.04, 0.06, 0.08, 0.1keV$, where $q=2$,$\beta_1=0.02$,$\beta_2=0.196$}
\end{figure}
\begin{figure}
\centerline{
    \includegraphics[scale=0.5,angle=-90]{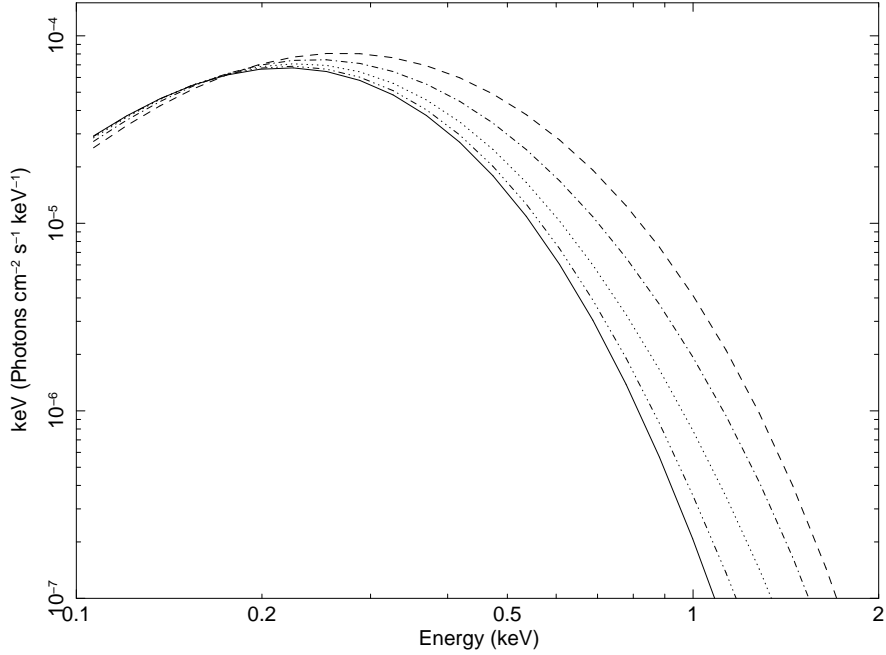}}
  \caption{The spectral shapes with different $q$, in which the dash, dash dot, dotted and solid lines correspond to $q=1, 2, 3, 4, 5$, and $kT_0=0.05keV$, $\beta_1=0.02$,$\beta_2=0.196$}
\end{figure}
\begin{figure}
\centerline{
    \includegraphics[scale=0.5,angle=-90]{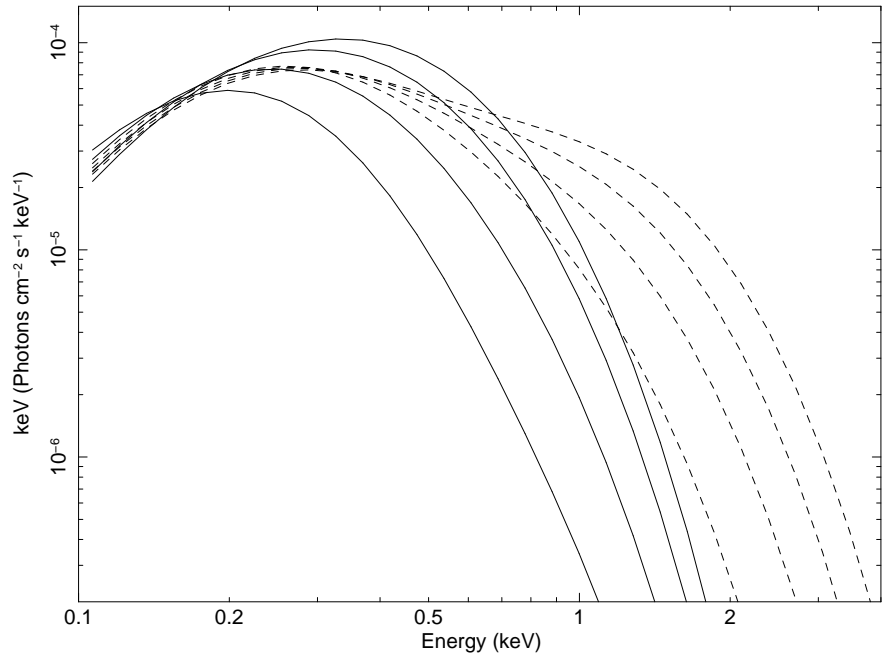}}
  \caption{The spectral shapes with different $\beta_1$ and $\beta_2$. Solid lines from down to up (above $0.5keV$) correspond to $\beta_1=0.002, 0.020, 0.059, 0.098$ when $\beta_2=0.196$.
  Dotted lines from down to up (above $0.5keV$) correspond to $\beta_2=0.391, 0.587, 0.783, 1$ when $\beta_1=0.02$. $kT_0=0.05keV$ and $q=2$ are used.}
\end{figure}

\begin{table*}
\begin{center}
\begin{tabular}{lcllc}
  \hline
  Source &  z & logM/M$_{\odot}$ &  L/L$_{Edd}$  &Observation ID\\
 \hline

 PG 0003+199  & 0.025 & 7.07 & 0.62  &0306870101\\
 PG 0844+349  & 0.064 & 7.66 & 0.41  &0554710101\\
 PG 0947+396  & 0.206 & 8.46 & 0.14  &0111290101\\
 PG 0953+414  & 0.239 & 8.52 & 0.58  &0111290201\\
 PG 1048+342  & 0.167 & 8.14 & 0.25  &0109080701\\
 PG 1116+215  & 0.177 & 8.41 & 0.74  &0111290401\\
 PG 1202+281  & 0.165 & 8.37 & 0.11  &0109080101\\
 PG 1211+143  & 0.080 & 7.81 & 1.14  &0112610101\\
 PG 1244+026  & 0.048 & 6.24 & 3.97  &0051760101\\
 PG 1307+085  & 0.155 & 8.50 & 0.24  &0110950401\\
 PG 1309+355  & 0.184 & 8.20 & 0.33  &0109080201\\
 PG 1322+659  & 0.168 & 7.74 & 0.81  &0109080301\\
 PG 1352+183  & 0.158 & 8.20 & 0.29  &0109080401\\
 PG 1402+261  & 0.164 & 7.76 & 1.24  &0109081001\\
 PG 1427+480  & 0.221 & 7.86 & 0.57  &0109080901\\
 PG 1440+356  & 0.077 & 7.28 & 1.07  &0107660201\\
 PG 1444+407  & 0.267 & 8.17 & 0.71  &0109080601\\
 PG 1501+106  & 0.036 & 8.23 & 0.12  &0070740301\\
 NGC 4051     & 0.002 & 6.13 & 0.31  &0157560101\\
 MRK 1044     & 0.016 & 6.23 & 1.25  &0112600301\\
 MRK 0359     & 0.017 & 6.23 & 1.79  &0112600601\\
 RE J1034+396 & 0.042 & 6.45 &0.28   &0561580201\\
 PKS 0558-504 & 0.137 & 7.65 &1.72   &0125110101\\
 MRK 0586     & 0.155 & 7.86 & 3.37  &0048740101\\
 TON S180     & 0.062 & 7.06 & 2.98  &0110890401\\

 \hline

\end{tabular}
\caption{Source properties of AGNs analyzed. Mass and luminosity are taken from \citet{mid07}}

\end{center}
\end{table*}

\begin{table*}
\begin{center}
\begin{tabular}{lccccccc}
\hline\hline
  source     &$n_H$ &$kT_0$ &$\beta_1$ &$\beta_2$ &$q$ &$\Gamma$ &reduced $\chi^2$(d.o.f.)\\
\hline\\
  PG0003+199 &$2.83$            &$0.07$      &$0.053$    &$0.770$     &$3.60$      &$1.92$         &1.294(973)\\
  PG0844+349 &$2.32$            &$0.05$      &$0.063$    &$0.571$     &$3.35$      &$0.97$         &0.995(297)\\
  PG0947+396 &$2.22$            &$0.03$      &$0.028$    &$1$         &$3.14$      &$1.97$         &0.946(866)\\
  PG0953+414 &$0.90$            &$0.01$      &$0.066$    &$1$         &$1.01$      &$2.32$         &0.966(871)\\
  PG1048+342 &$1.20$            &$0.08$      &$0.002$    &$0.461$     &$1.82$      &$1.68$         &0.927(412)\\
  PG1116+215 &$1.78$            &$0.04$      &$0.027$    &$1$         &$3.25$      &$2.31$         &0.951(928)\\
  PG1202+281 &$1.03$            &$0.04$      &$0.074$    &$1$         &$3.68$      &$1.65$         &1.003(598)\\
  PG1211+143 &$3.26$            &$0.02$      &$0.007$    &$1$         &$2.85$      &$2.06$         &1.164(1028)\\
  PG1244+026 &$7.68$            &$0.04$      &$0.098$    &$0.493$     &$1.00$      &$2.42$         &1.234(208)\\
  PG1307+085 &$10.51$           &$0.01$      &$0.046$    &$0.561$     &$4.11$      &$1.72$         &1.545(230)\\
  PG1309+355 &$3.63$            &$0.01$      &$0.026$    &$0.767$     &$2.12$      &$1.75$         &1.147(284)\\
  PG1322+695 &$1.36$            &$0.04$      &$0.025$    &$0.489$     &$2.58$      &$2.15$         &1.058(450)\\
  PG1352+183 &$2.08$            &$0.04$      &$0.016$    &$0.622$     &$2.64$      &$2.02$         &1.104(435)\\
  PG1402+261 &$0.25$            &$0.04$      &$0.047$    &$1$         &$3.27$      &$2.12$         &0.972(667)\\
  PG1427+480 &$3.91$            &$0.01$      &$0.045$    &$1$         &$4.25$      &$2.24$         &1.270(494)\\
  PG1440+356 &$3.46$            &$0.04$      &$0.080$    &$0.366$     &$1.56$      &$2.46$         &1.006(680)\\
  PG1444+407 &$2.07$            &$0.03$      &$0.006$    &$0.678$     &$2.70$      &$2.31$         &1.122(297)\\
  PG1501+106 &$1.16$            &$0.05$      &$0.055$    &$1$         &$3.54$      &$1.80$         &0.974(818)\\
  NGC4051    &$1.04$            &$0.10$      &$0.035$    &$0.362$     &$2.95$      &$1.25$         &2.051(1121)\\
  Mrk1044    &$3.44$            &$0.08$      &$0.062$    &$0.196$     &$5.00$      &$2.02$         &1.387(185)\\
  Mrk0359    &$7.66$            &$0.04$      &$0.031$    &$0.608$     &$2.79$      &$1.92$         &1.004(877)\\
  REJ1034+396&$4.90$            &$0.04$      &$0.098$    &$0.290$     &$1.42$      &$2.50$         &1.196(746)\\
  PKS0558-504&$3.60$            &$0.04$      &$0.027$    &$0.608$     &$2.92$      &$2.25$         &1.108(712)\\
  Mrk0586    &$6.89$            &$0.04$      &$0.090$    &$0.312$     &$1.09$      &$2.35$         &1.212(776)\\
  TON S180   &$4.09$            &$0.04$      &$0.077$    &$0.418$     &$1.57$      &$2.39$         &1.272(685)\\
\hline
\end{tabular}
\caption{The fitting results of the sources. $\Gamma$ is the hard X-ray spectrum index, $kT_0$ is the temperature of inner stable radius of accretion disk, in which $n_H$ is in unit of $10^{20}cm^{-2}$ and $kT_0$ is in unit of $keV$ ¡£}
\end{center}
\end{table*}

\begin{table*}
\begin{center}
\begin{tabular}{cccc}
\hline\hline
  source     &$L_{soft}(\times10^{44})$  &$\xi$  &$ratio(\%)$ \\
\hline\\
  PG0003+199$^a$ &$0.55$  &$13.00$ &$4.34$  \\
  PG0844+349$^b$ &$3.78$  &$4.13 $ &$10.90$  \\
  PG0947+396$^b$ &$2.55$  &$6.14 $ &$27.50$  \\
  PG0953+414$^b$ &$7.77$  &$2.96 $ &$8.69$  \\
  PG1048+342$^b$ &$1.07$  &$48.27$ &$6.03$  \\
  PG1116+215$^b$ &$5.00$  &$3.87 $ &$6.21$  \\
  PG1202+281$^b$ &$2.22$  &$6.65 $ &$33.89$  \\
  PG1211+143$^b$ &$0.16$  &$19.00$ &$0.82 $  \\
  PG1244+026$^b$ &$0.36$  &$15.04$ &$0.66 $  \\
  PG1307+085$^b$ &$0.36$  &$5.33 $ &$3.39 $  \\
  PG1309+355$^b$ &$0.46$  &$6.33 $ &$2.30$  \\
  PG1322+695$^b$ &$2.21$  &$6.00 $ &$4.80$  \\
  PG1352+183$^b$ &$1.85$  &$7.32$  &$12.36$  \\
  PG1402+261$^b$ &$2.70$  &$5.25$  &$3.30$  \\
  PG1427+480$^b$ &$2.00$  &$2.22$  &$1.93$  \\
  PG1440+356$^b$ &$1.00$  &$9.35$  &$3.02$  \\
  PG1444+407$^b$ &$2.60$  &$4.96$  &$4.71$  \\
  PG1501+106$^b$ &$0.57$  &$20.07$ &$26.71$    \\
  NGC4051$^a$    &$0.015$ &$146.80$ &$2.82$  \\
  Mrk1044$^a$    &$0.16$  &$120.84$ &$7.69$  \\
  Mrk0359$^c$    &$0.07$  &$46.08$ &$0.86$   \\
  REJ1034+396    &-       &-       &-        \\
  PKS0558-504$^d$&$68.1$ &$0.59$ &$4.24$    \\
  Mrk0586$^c$    &$3.55$ &$4.43$ &$1.29$     \\
  TON S180$^a$   &$68.1$  &$0.53 $ &$2.46$  \\
\hline
\end{tabular}
\caption{The estimated parameters by the model. $L_{soft}$ is the observed luminosity. a: luminosity are taken from ~\citet{gru10}, b: luminosity are taken from~\citet{pic05}£¬c: luminosity are taken from NED(http://ned.ipac.caltech.edu/), d: luminosity are taken from~\citet{mah10}£¬and $ratio$ is the magnetic energy conversion rate.}
\end{center}
\end{table*}

\begin{figure*}
\includegraphics[scale=0.2,angle=-90]{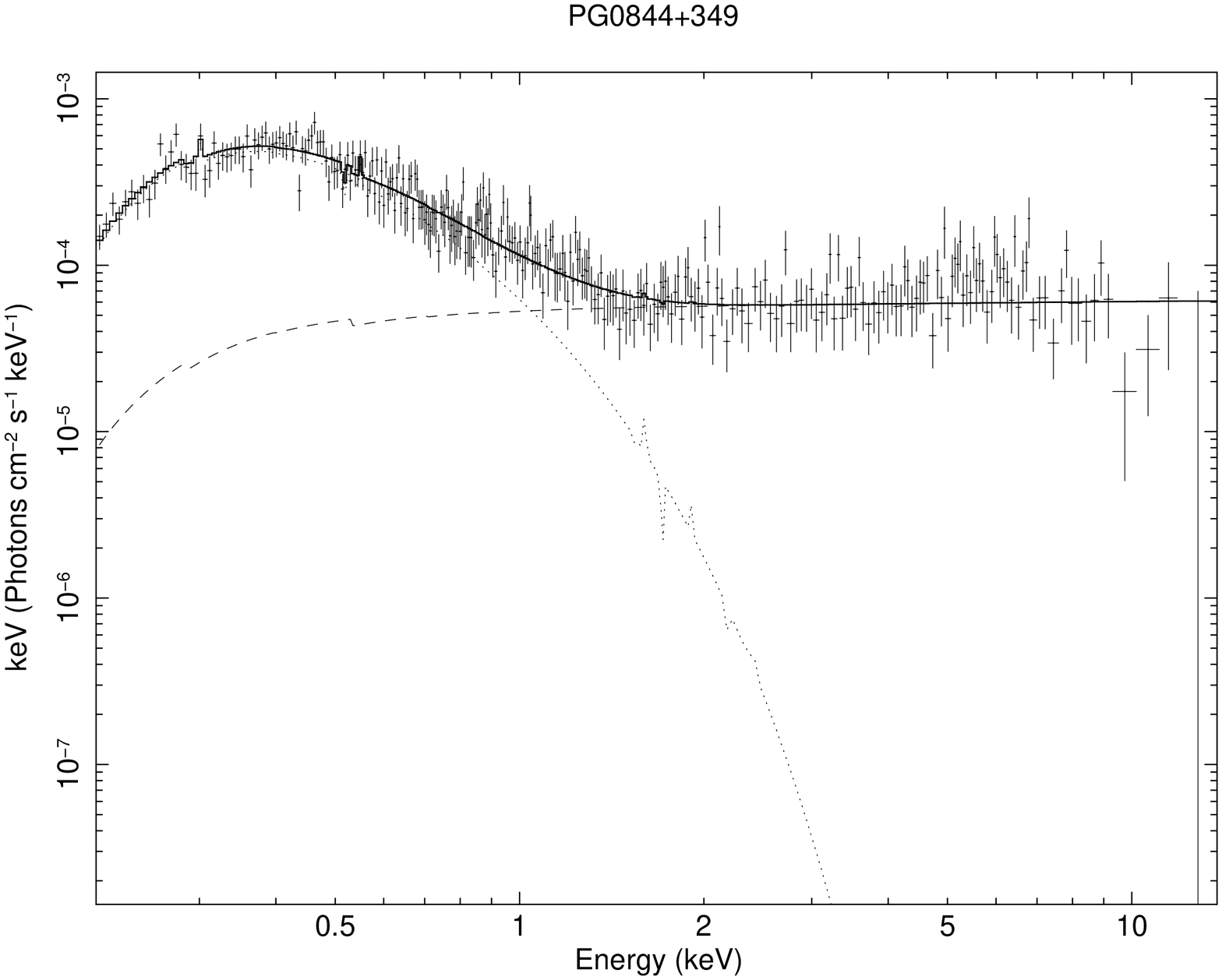}
\includegraphics[scale=0.2,angle=-90]{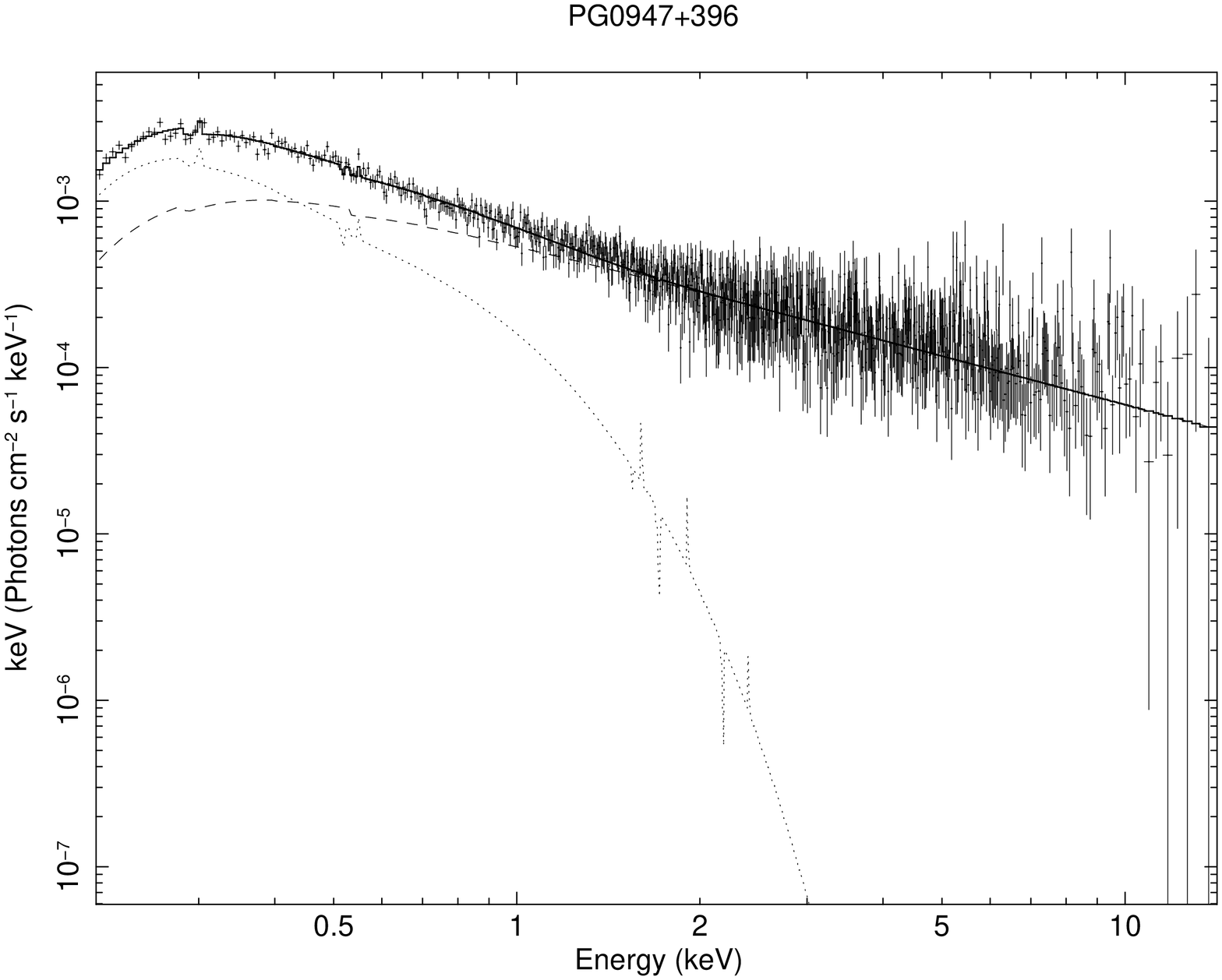}
\includegraphics[scale=0.2,angle=-90]{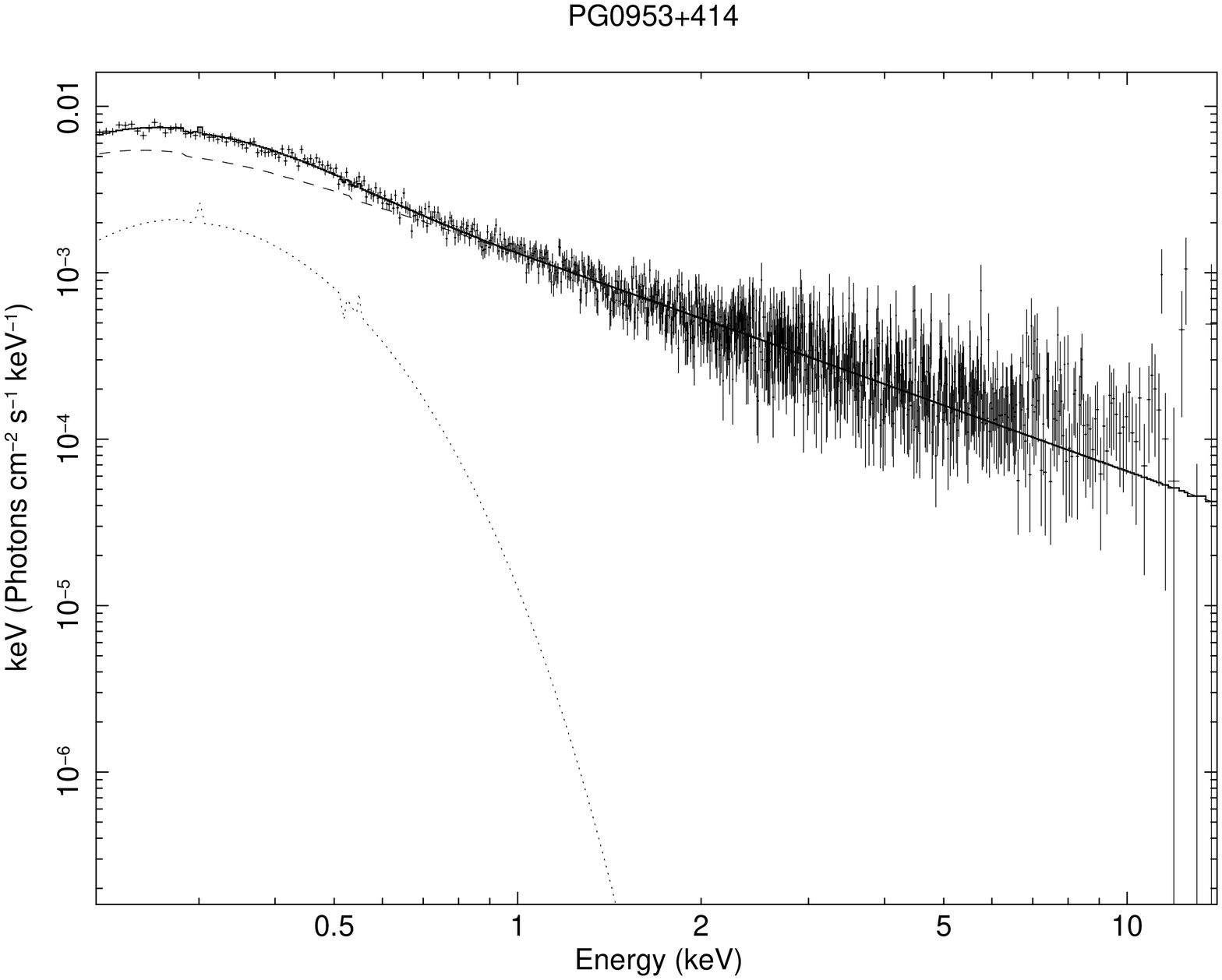}\\
\includegraphics[scale=0.2,angle=-90]{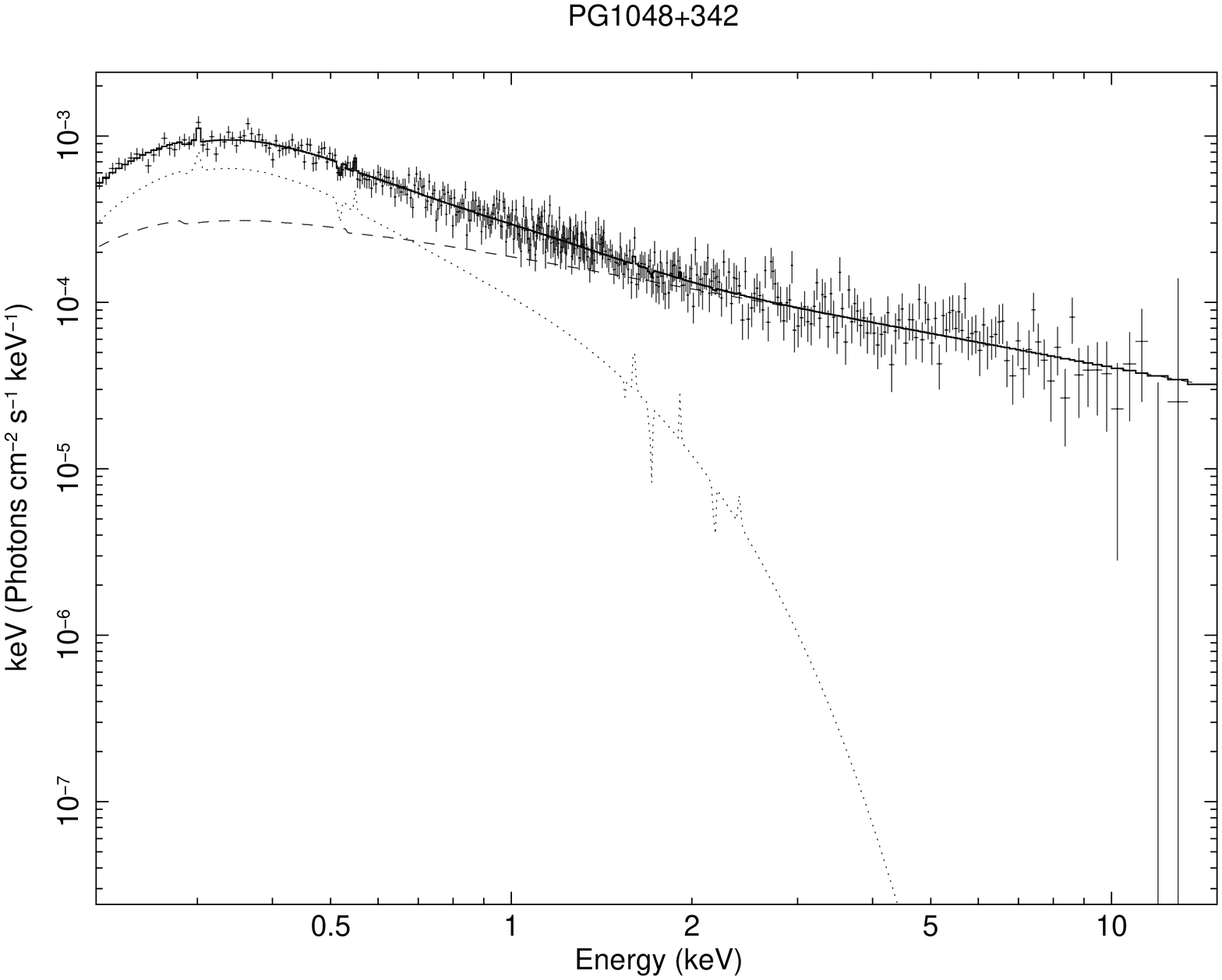}
\includegraphics[scale=0.2,angle=-90]{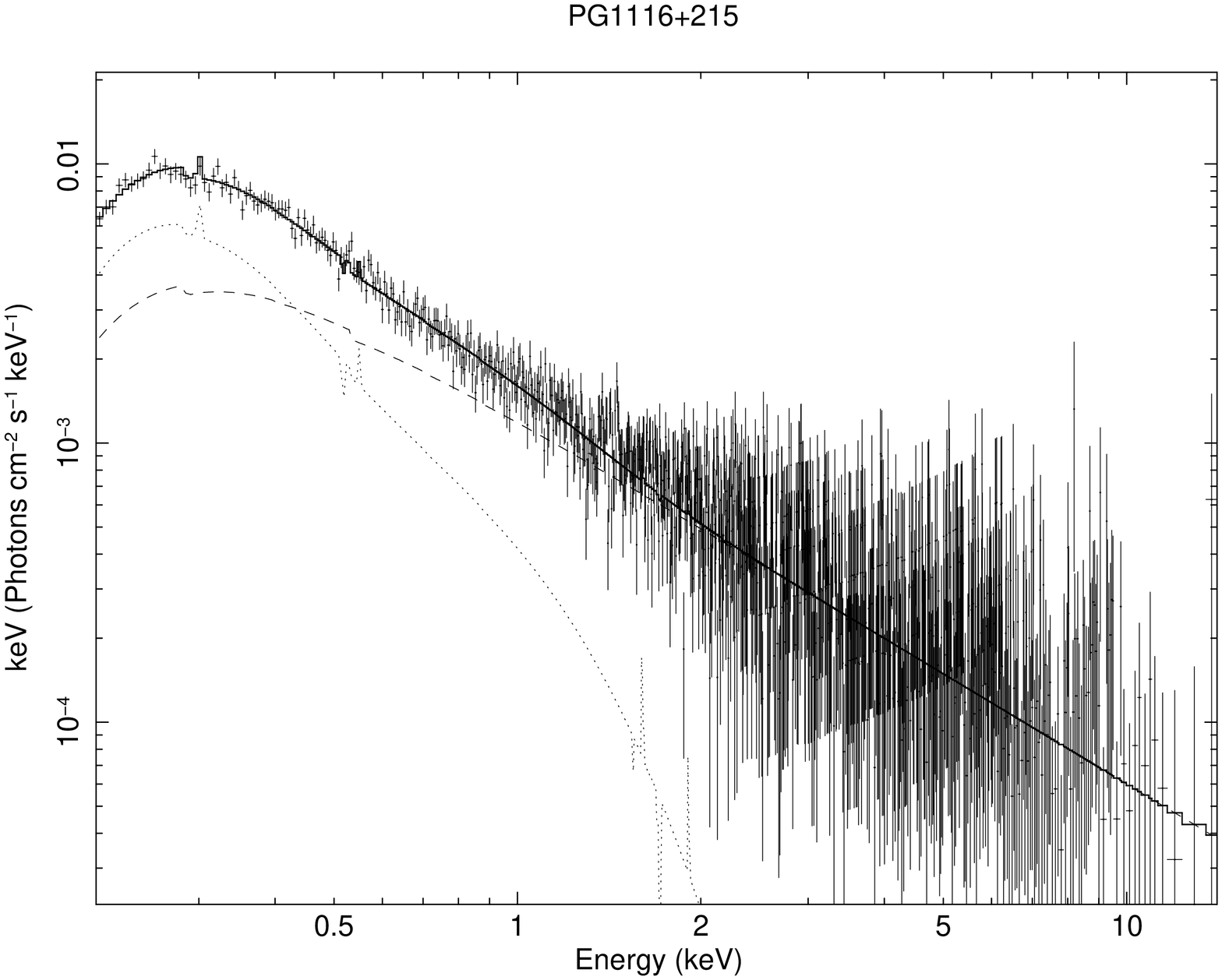}
\includegraphics[scale=0.2,angle=-90]{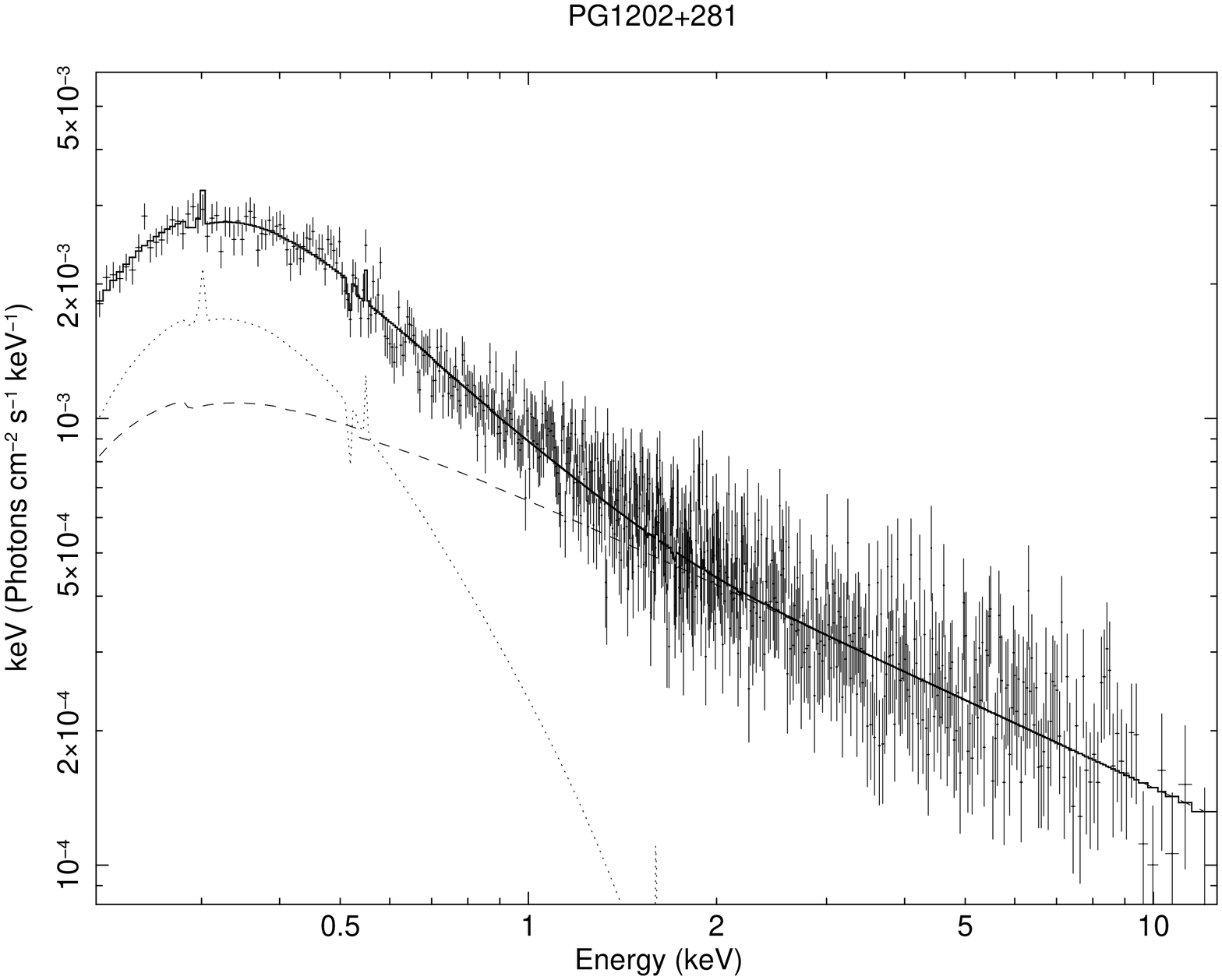}\\
\includegraphics[scale=0.2,angle=-90]{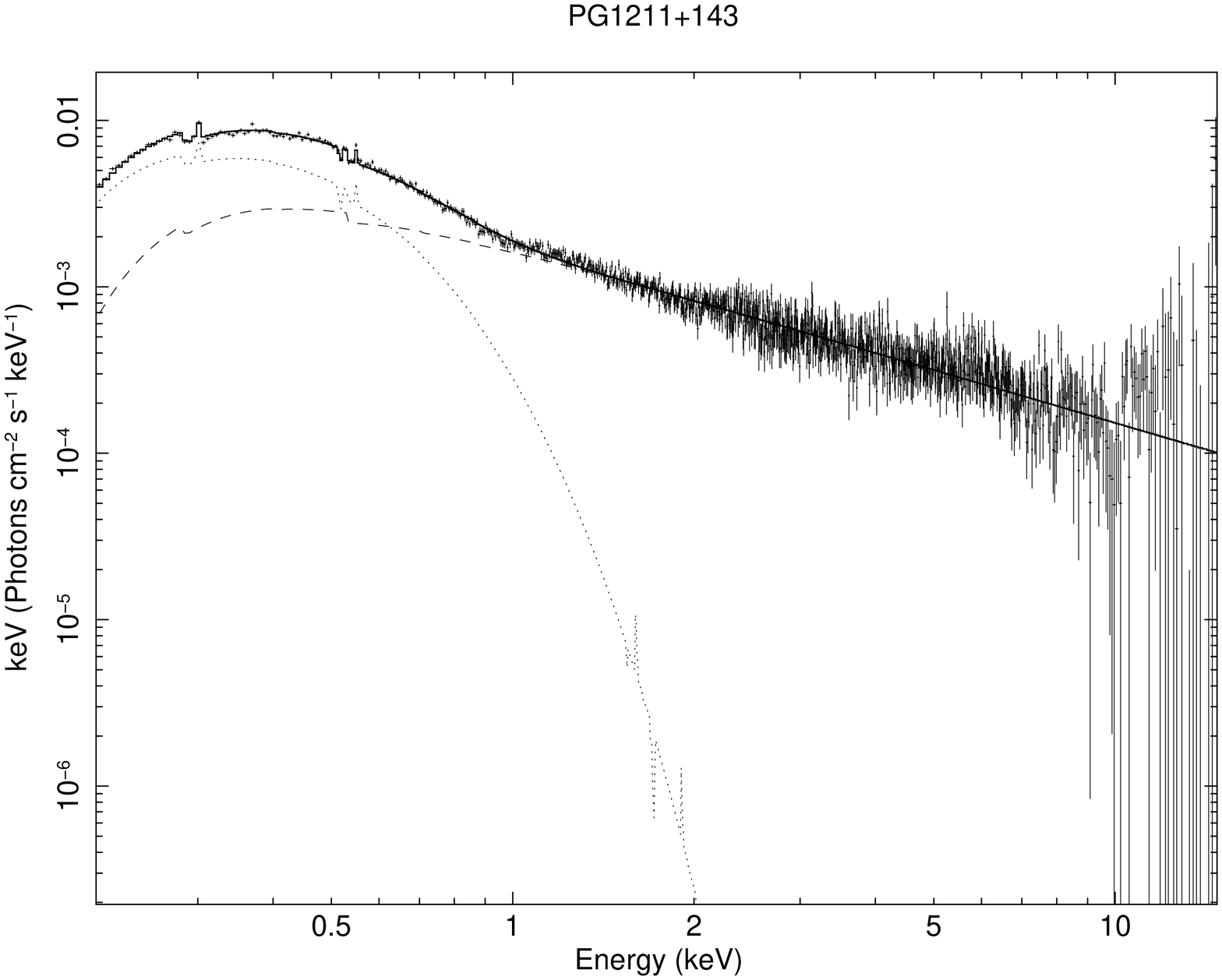}
\includegraphics[scale=0.2,angle=-90]{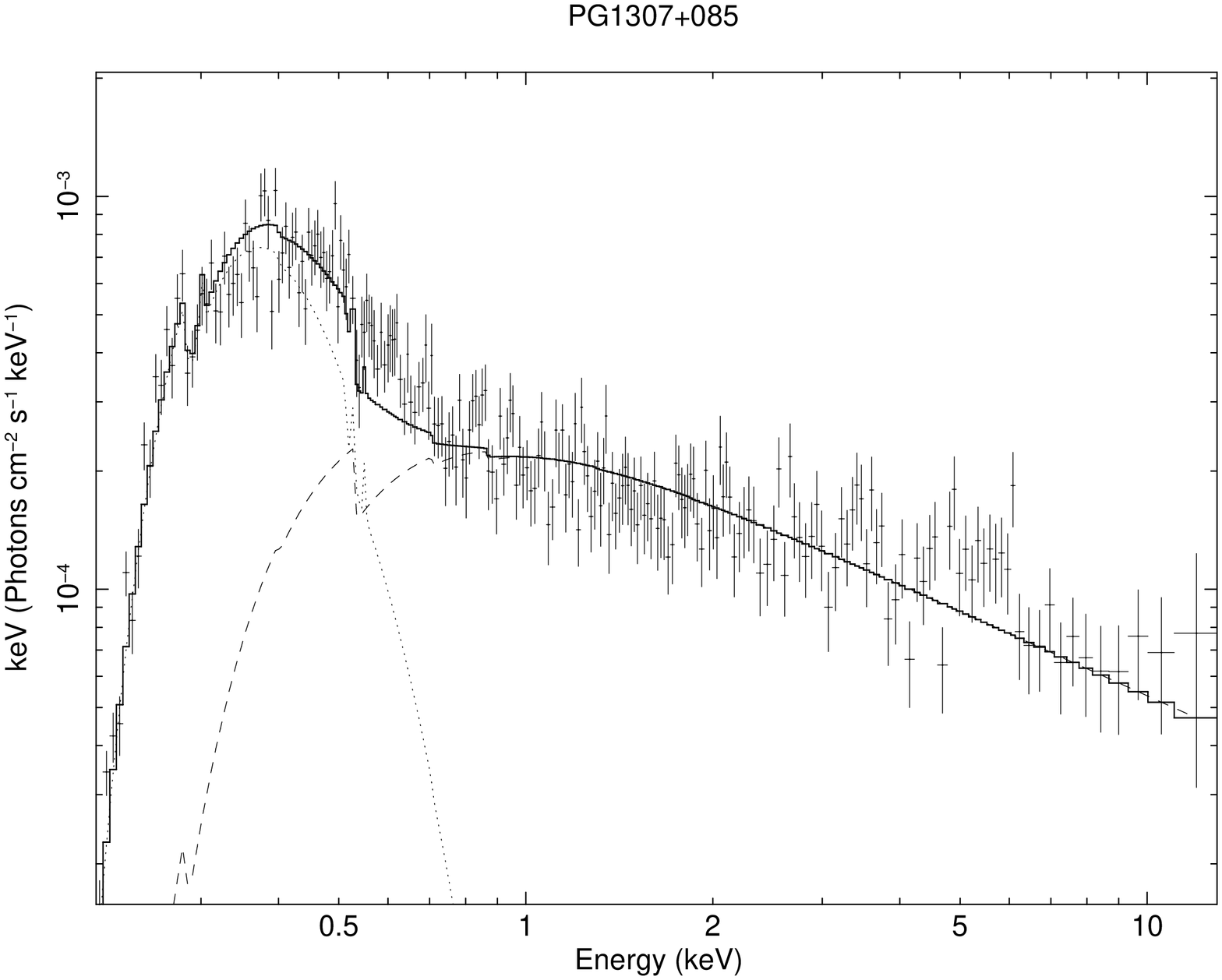}
\includegraphics[scale=0.2,angle=-90]{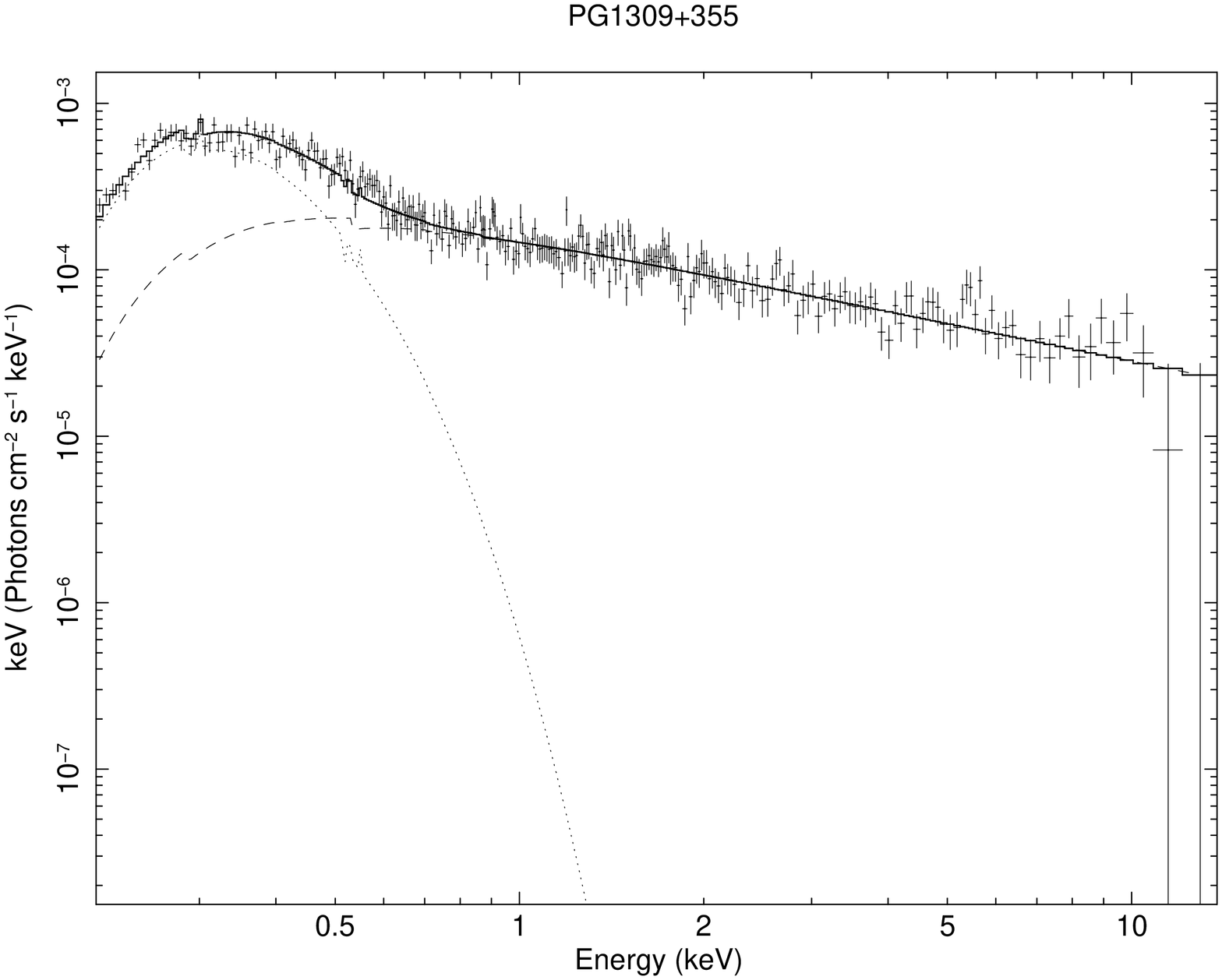}\\
\includegraphics[scale=0.2,angle=-90]{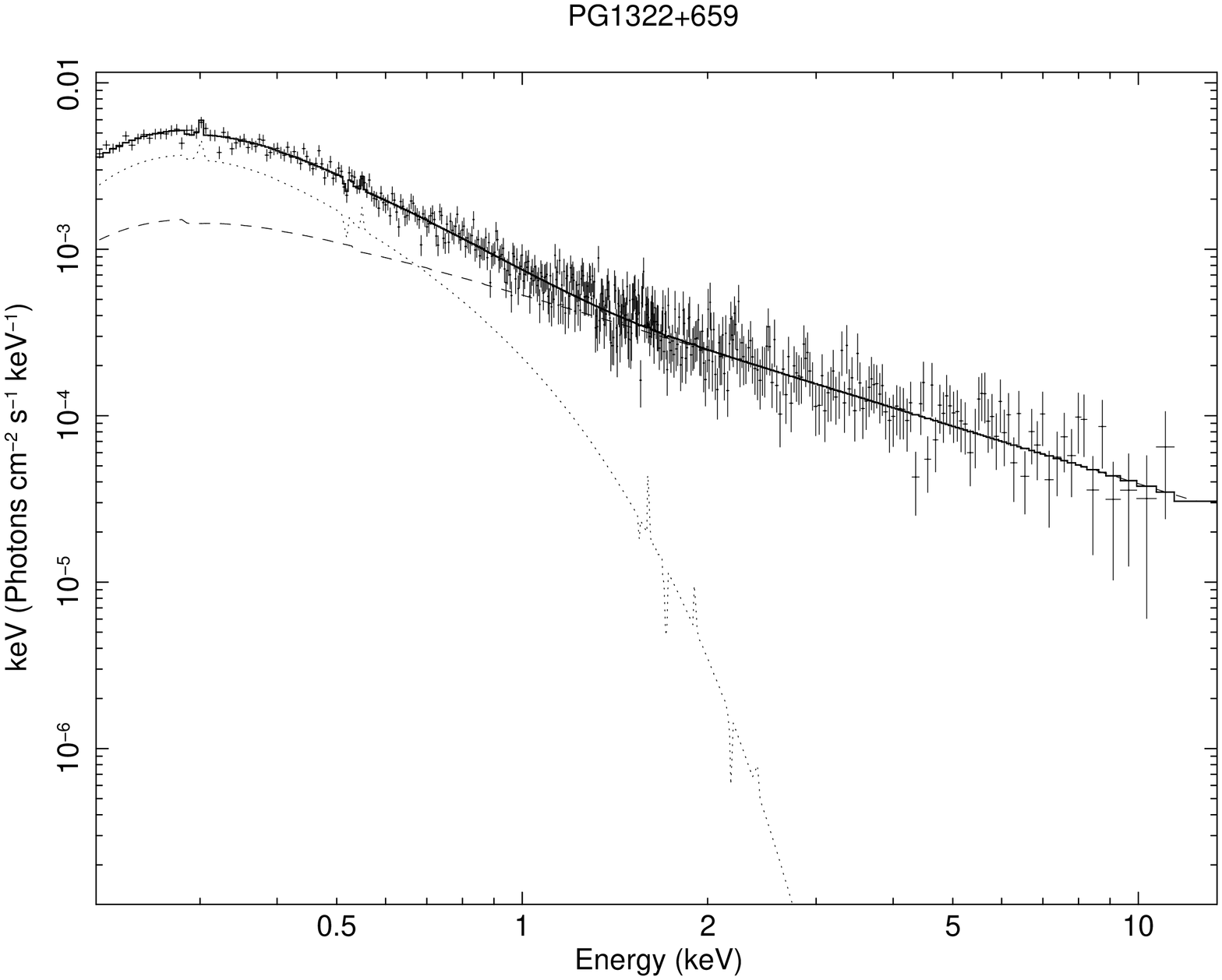}
\includegraphics[scale=0.2,angle=-90]{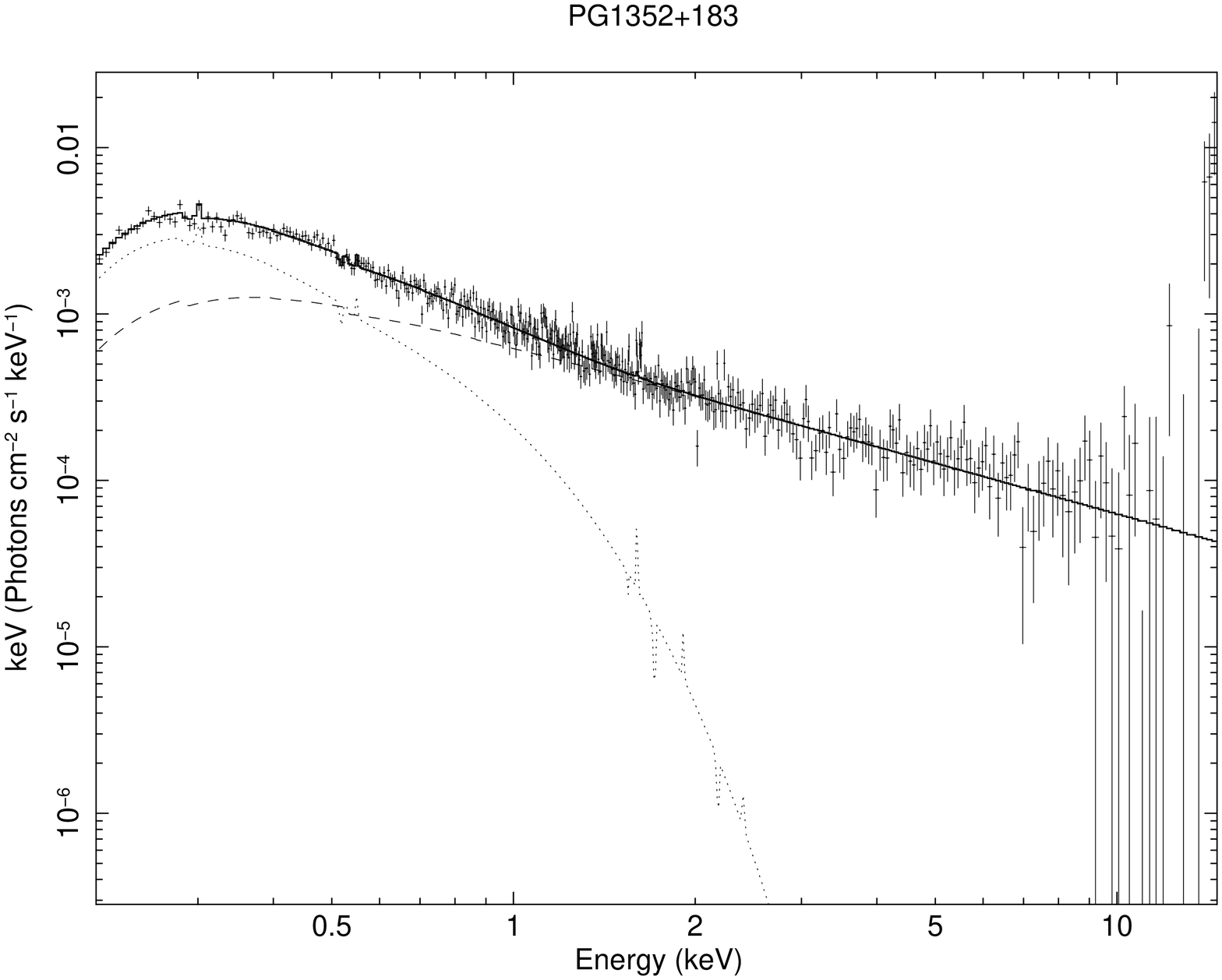}
\includegraphics[scale=0.2,angle=-90]{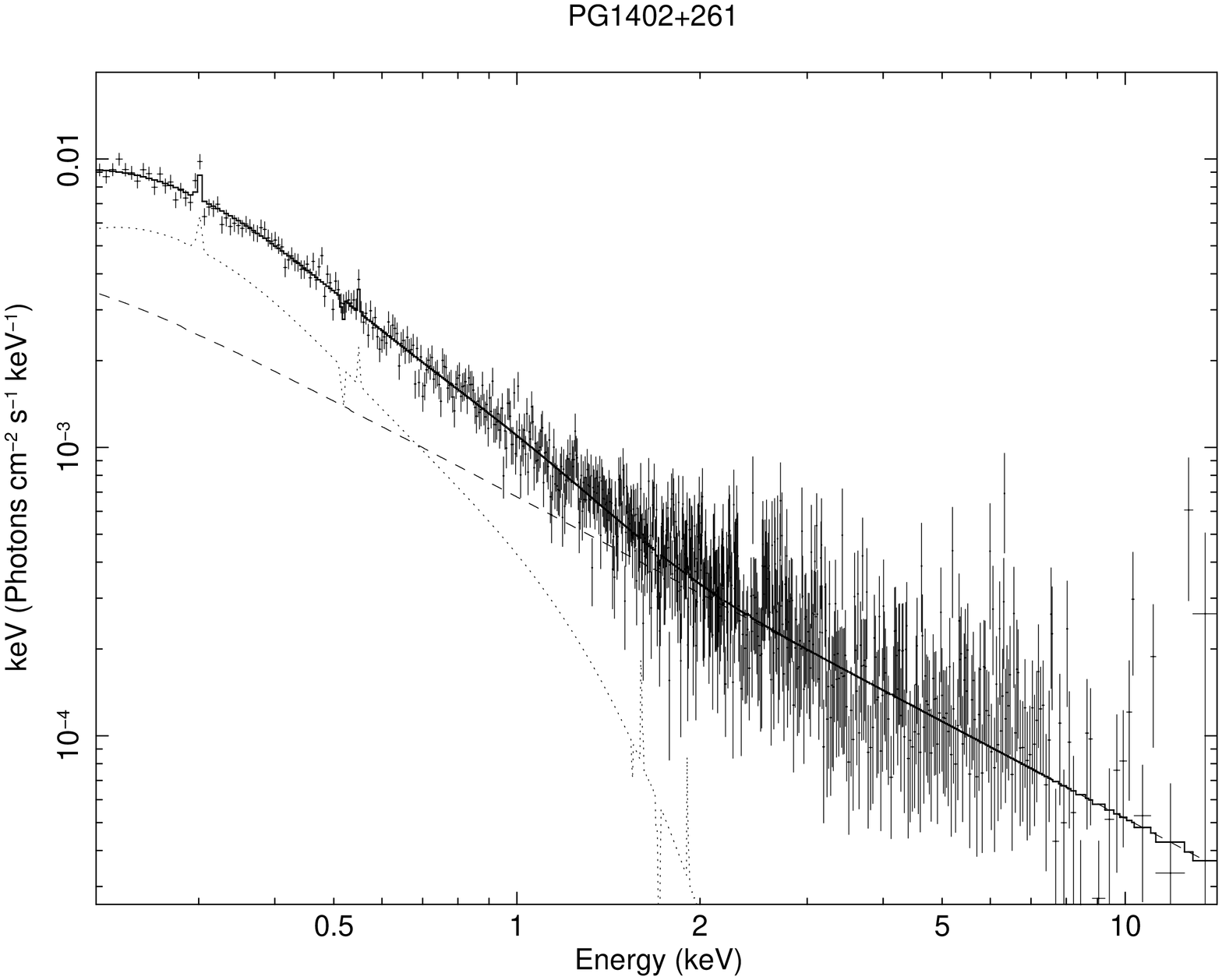}\\
\includegraphics[scale=0.2,angle=-90]{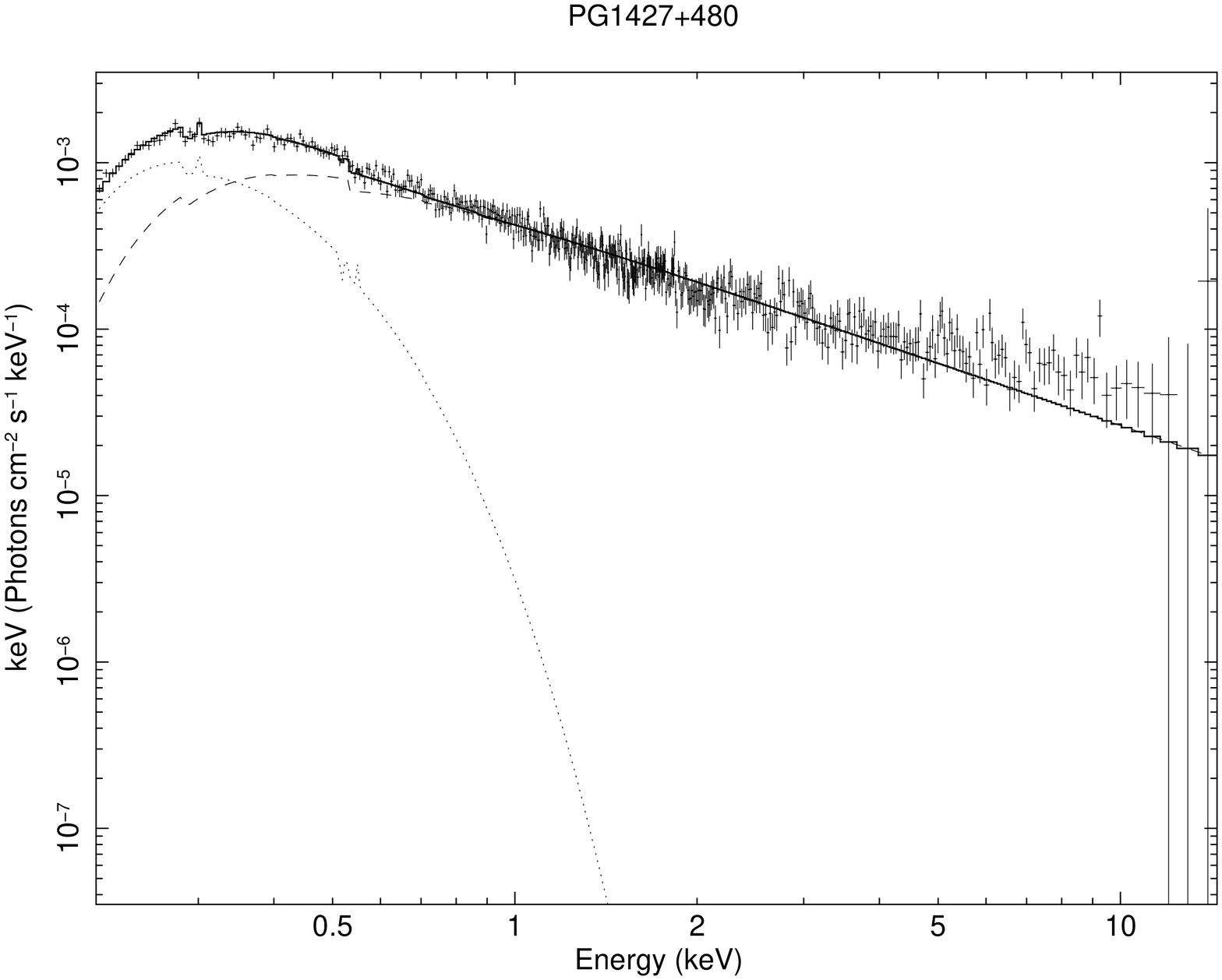}
\includegraphics[scale=0.2,angle=-90]{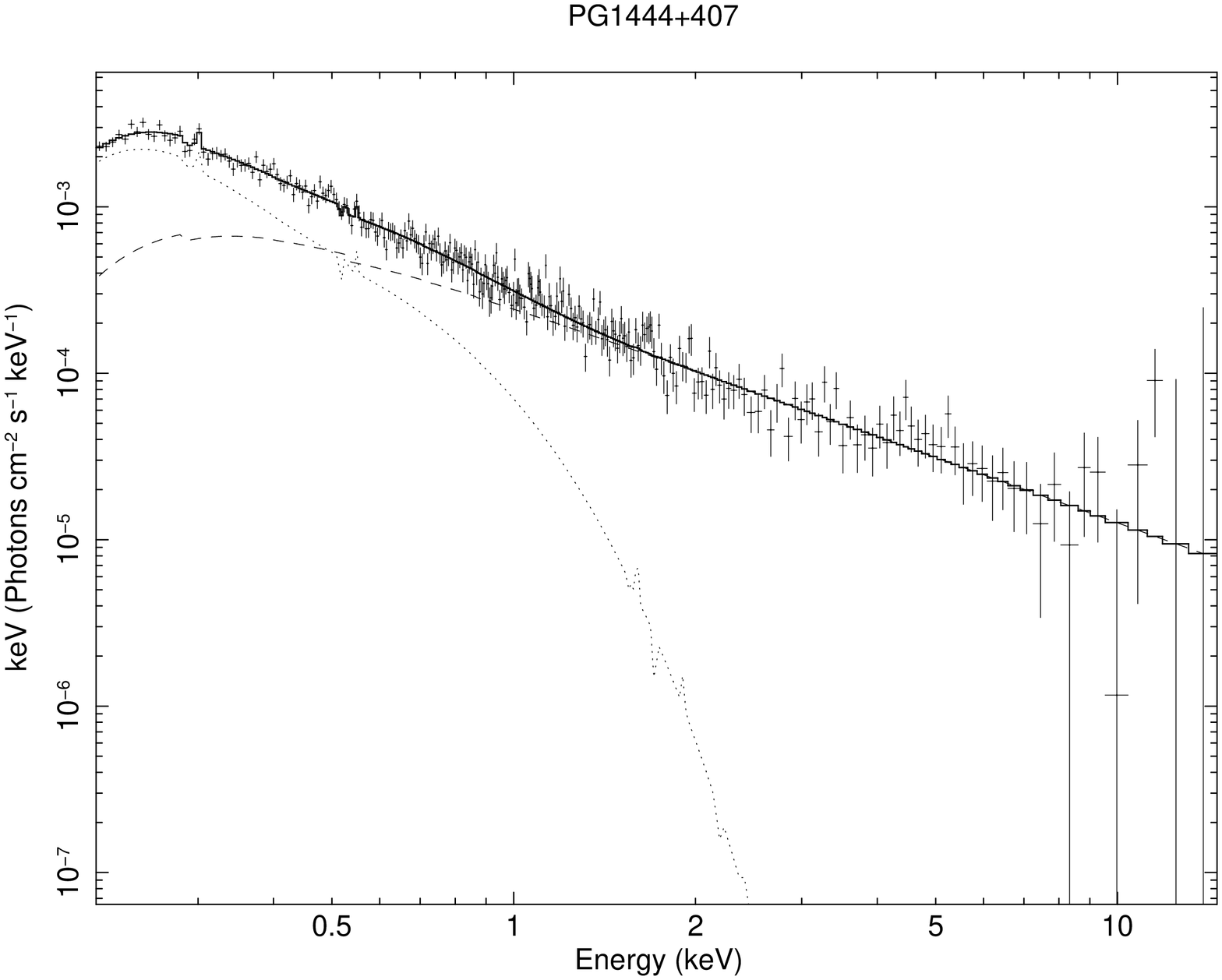}
\includegraphics[scale=0.2,angle=-90]{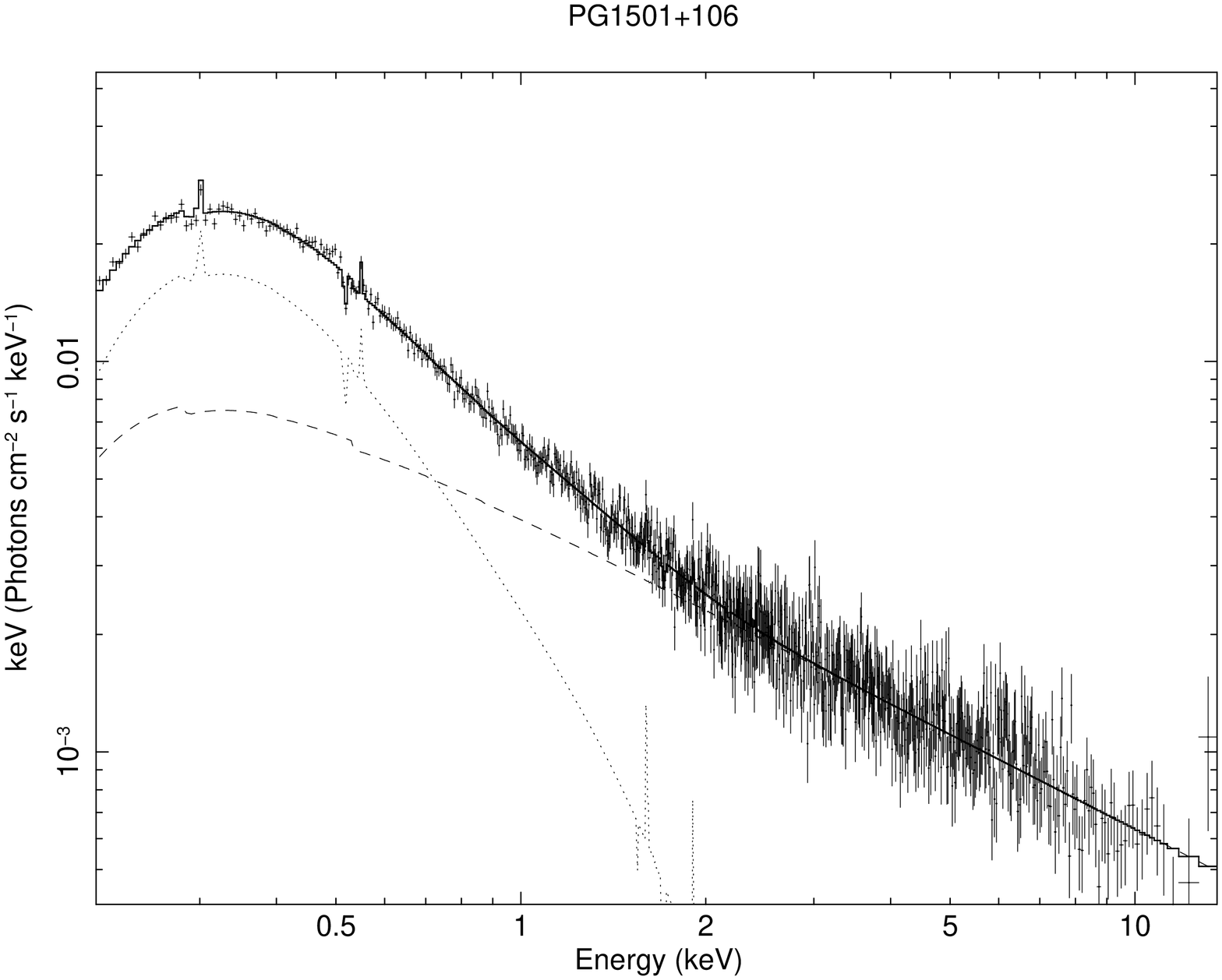}\\
\includegraphics[scale=0.2,angle=-90]{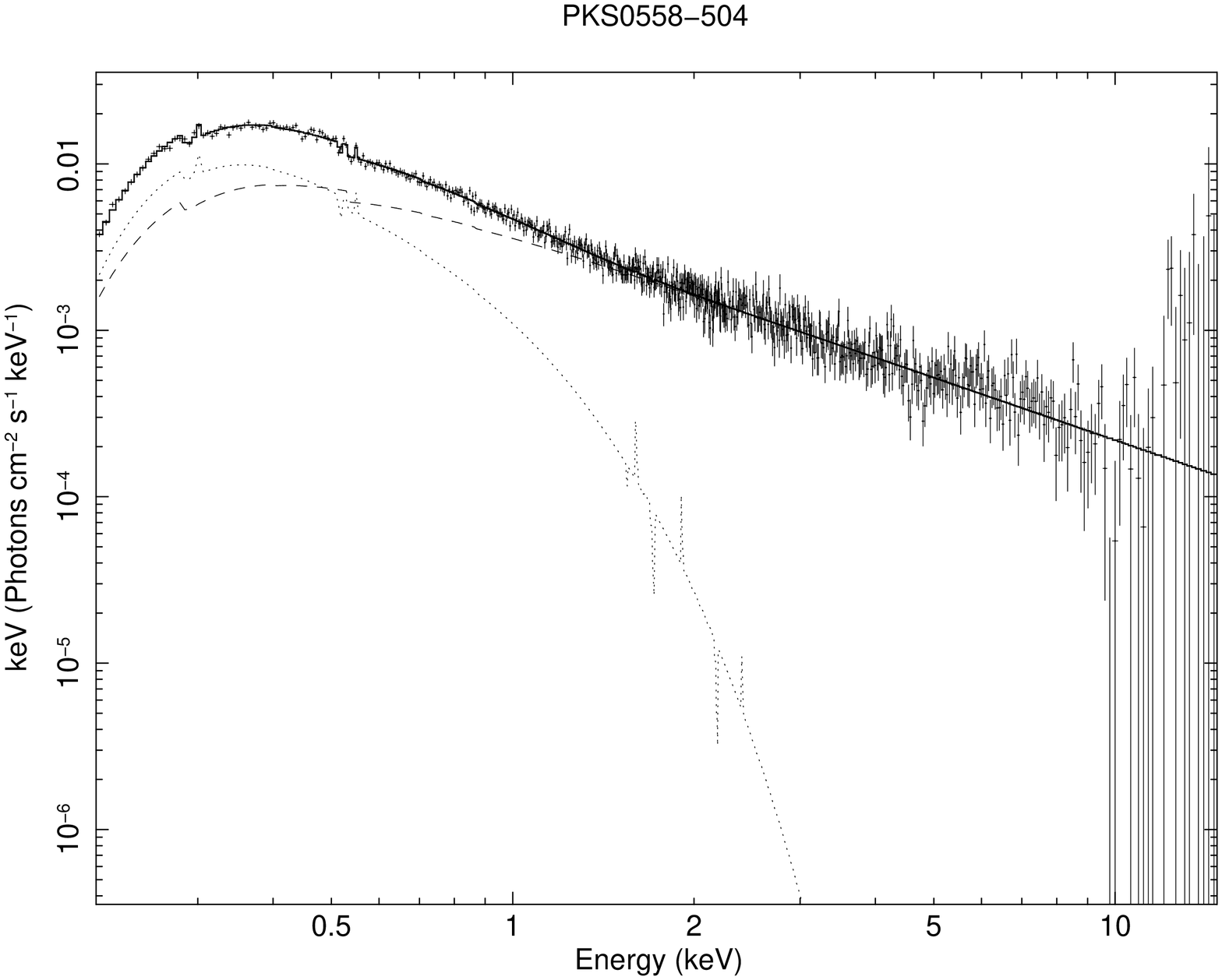}
\caption[spectrum]{\label{Photoz}The fitting spectra. The black crosses are the data points, the solid line is the best fit to data. The dot line shows the inverse Compton scattering component from disc, and the dash line is the power law component from corona.}
\end{figure*}

\begin{figure*}
\includegraphics[scale=0.2,angle=-90]{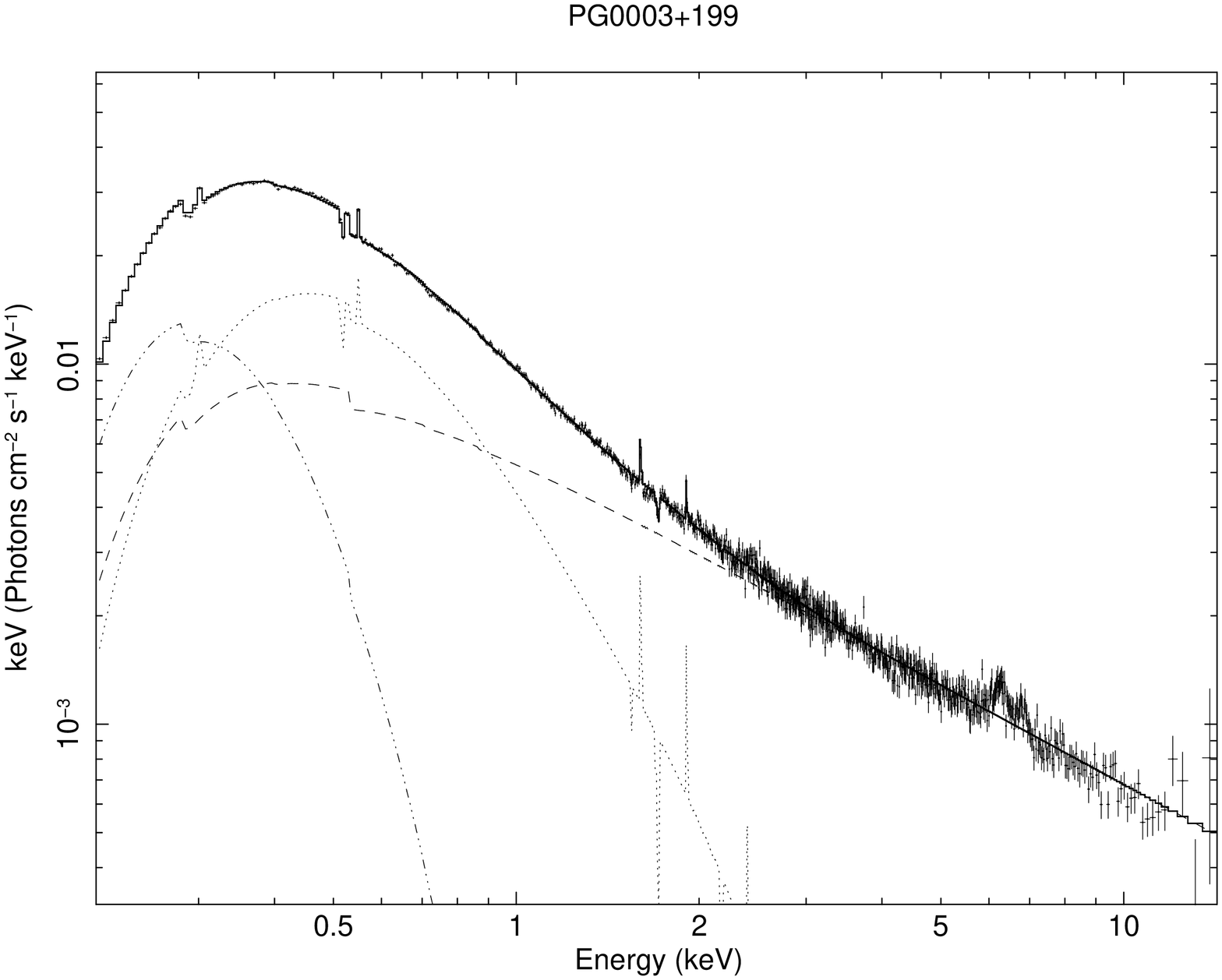}
\includegraphics[scale=0.2,angle=-90]{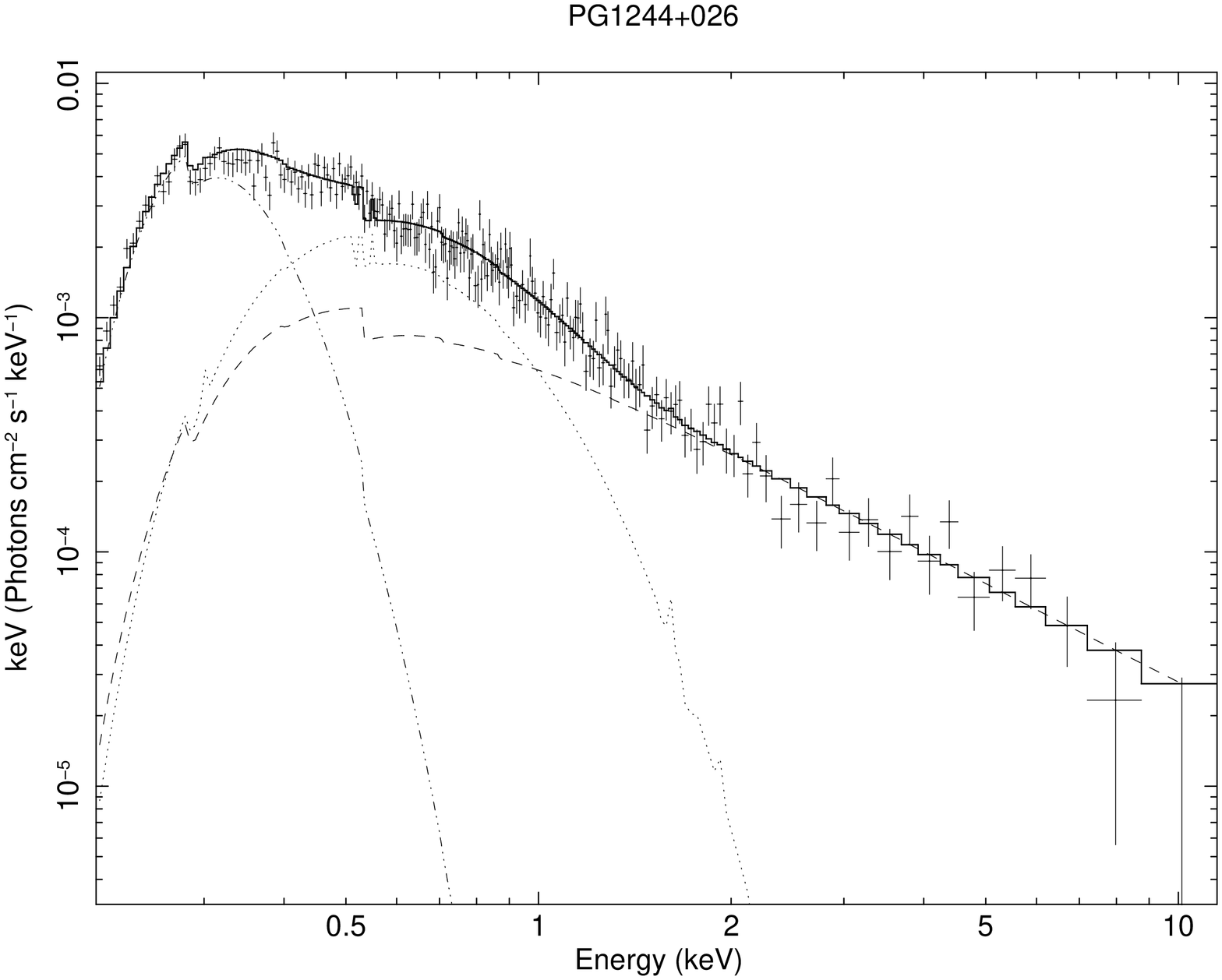}
\includegraphics[scale=0.2,angle=-90]{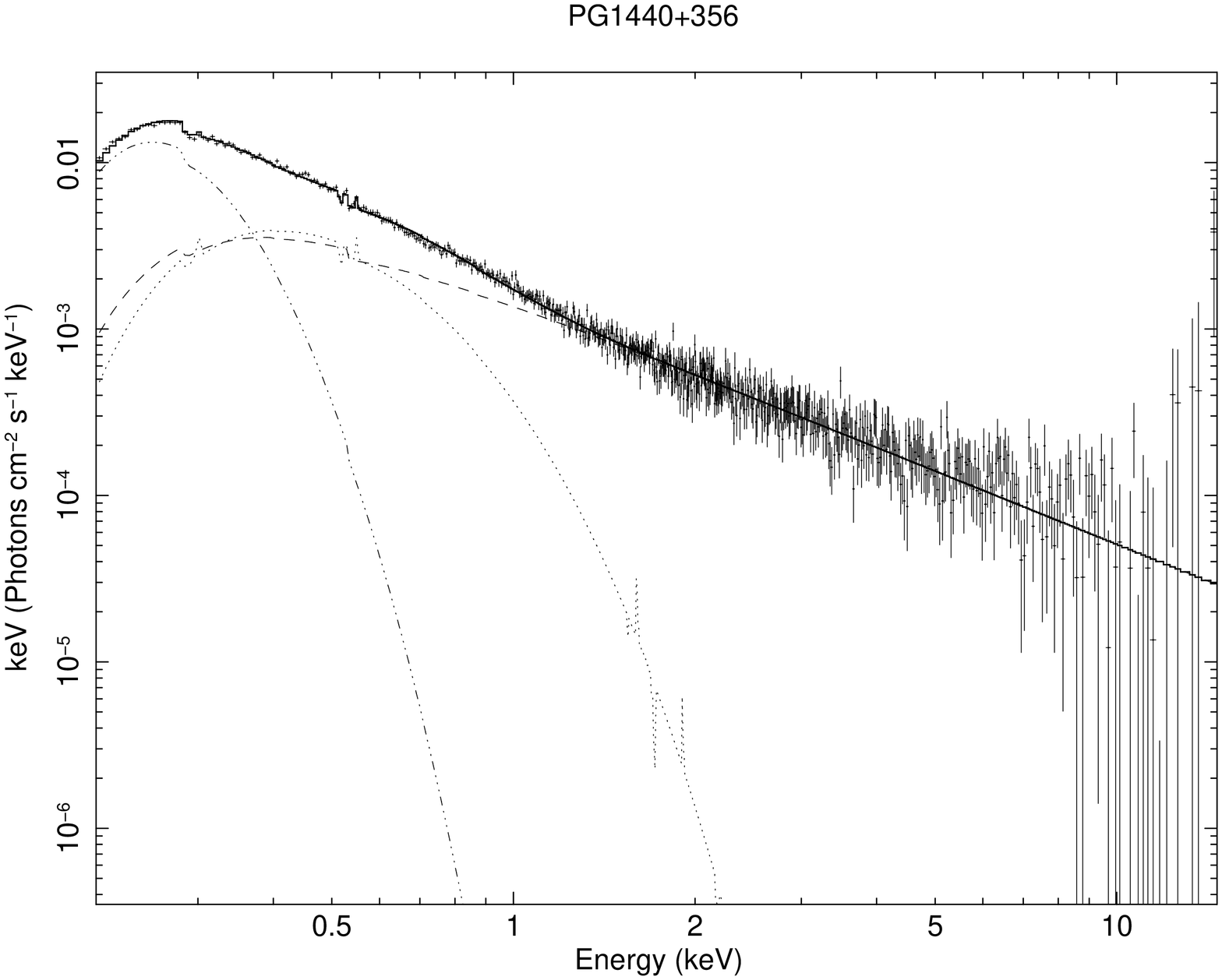}\\
\includegraphics[scale=0.2,angle=-90]{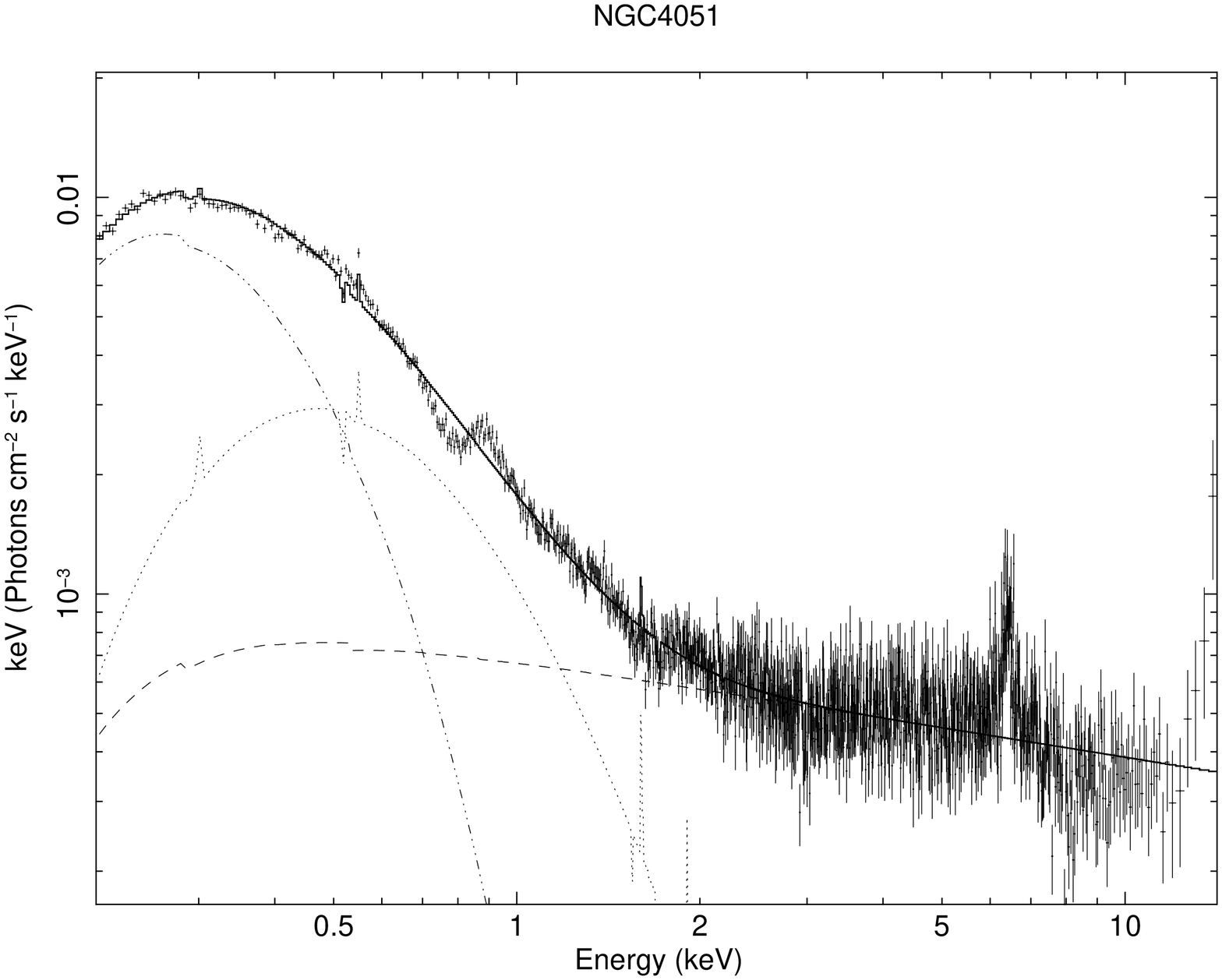}
\includegraphics[scale=0.2,angle=-90]{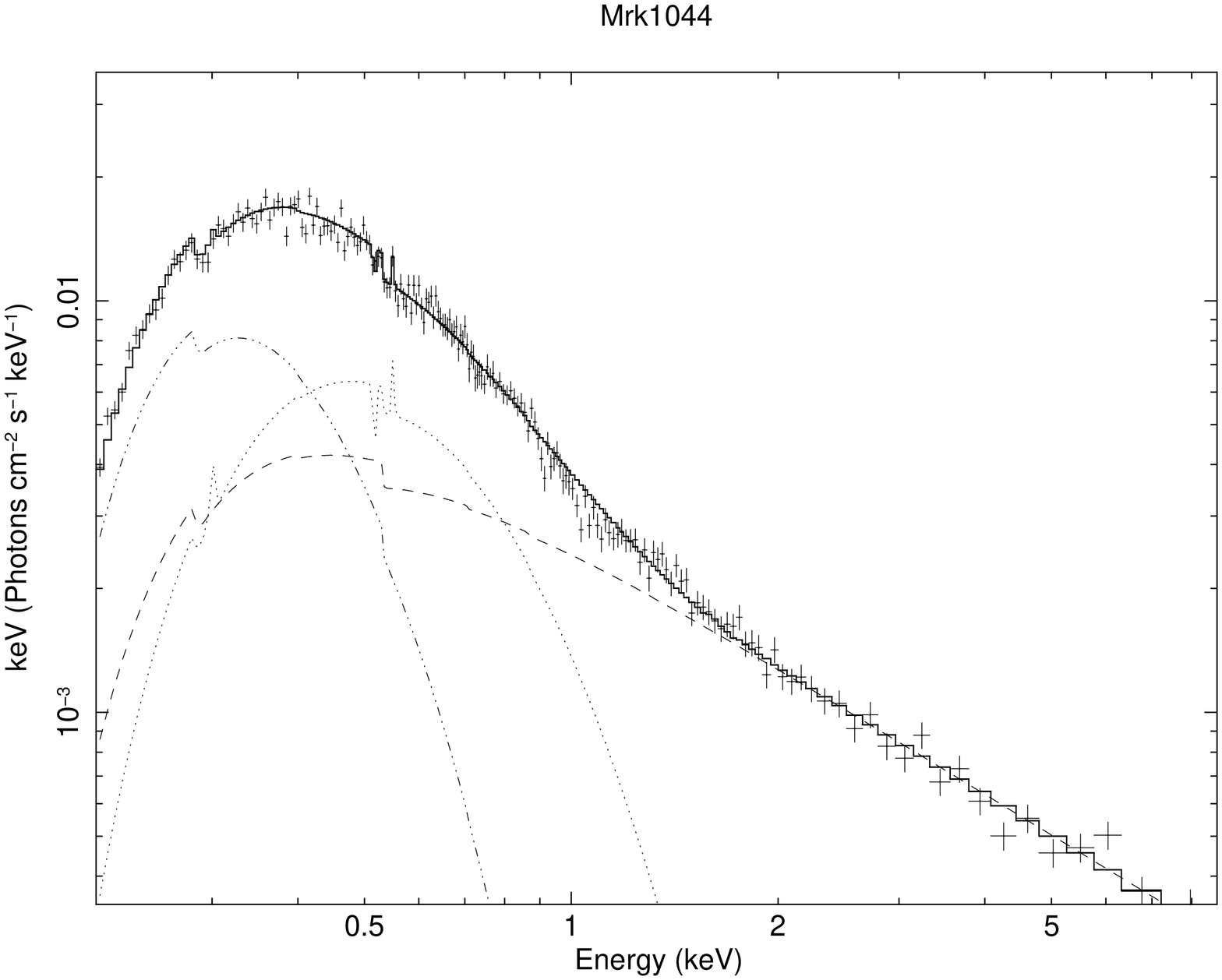}
\includegraphics[scale=0.2,angle=-90]{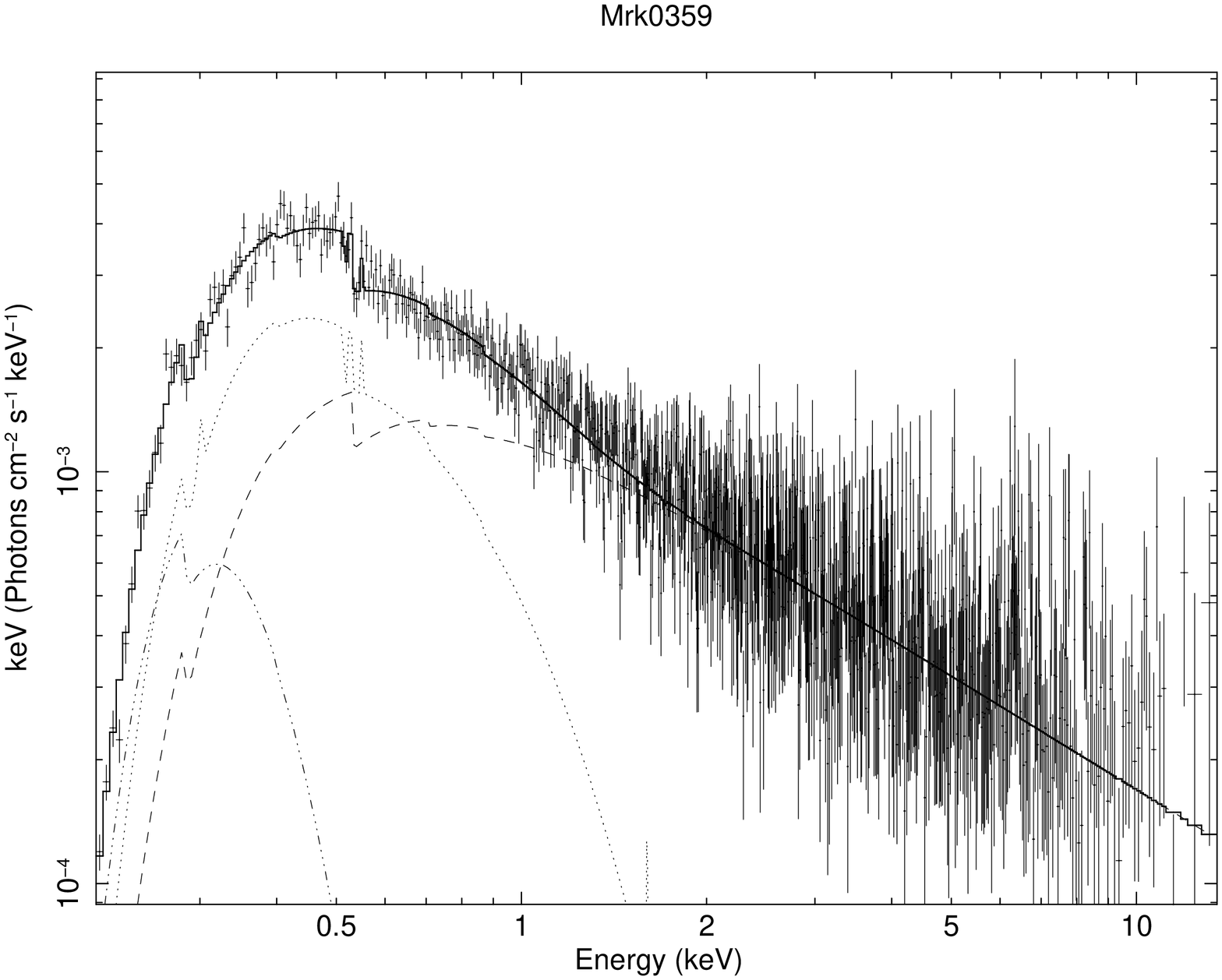}\\
\includegraphics[scale=0.2,angle=-90]{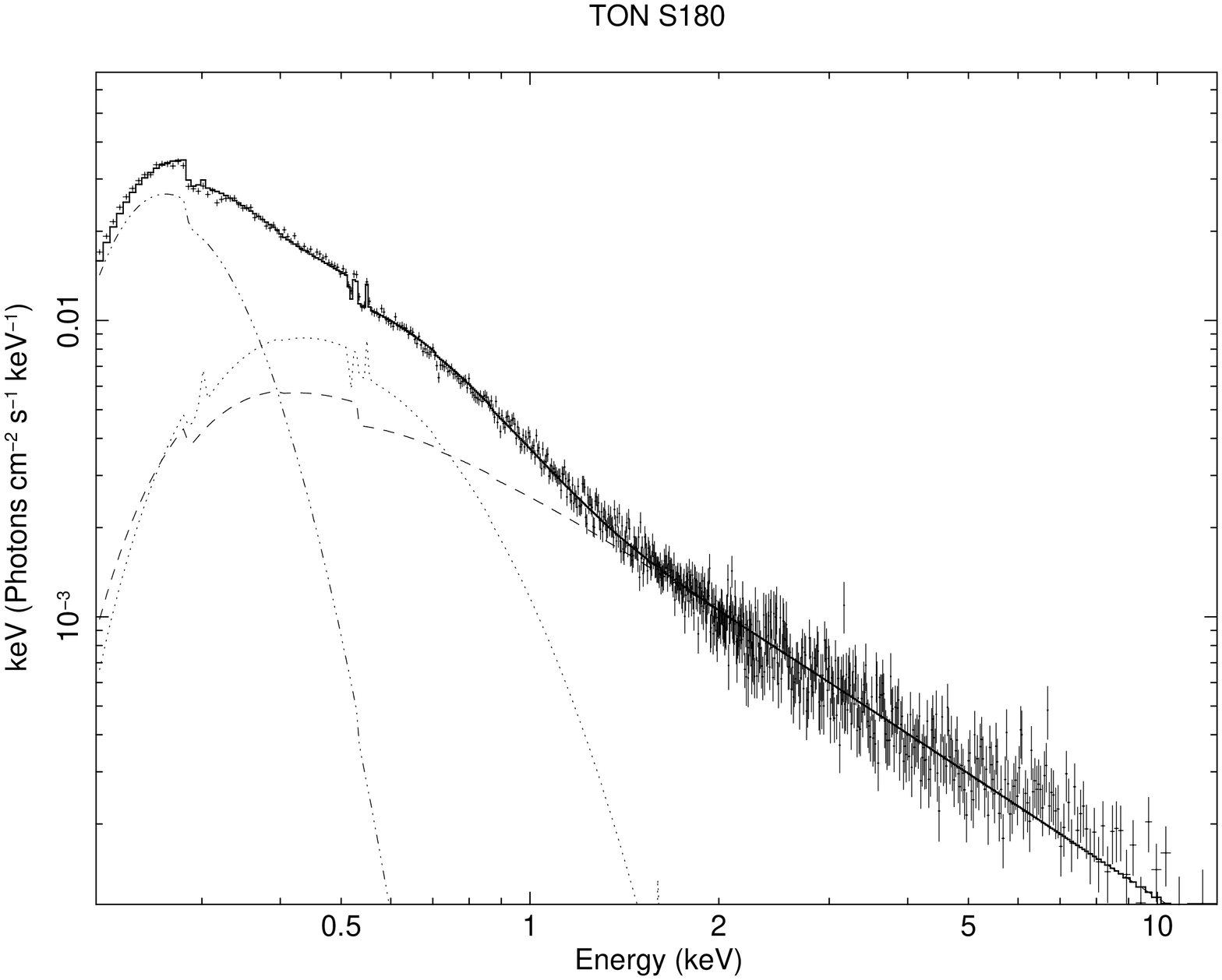}
\includegraphics[scale=0.2,angle=-90]{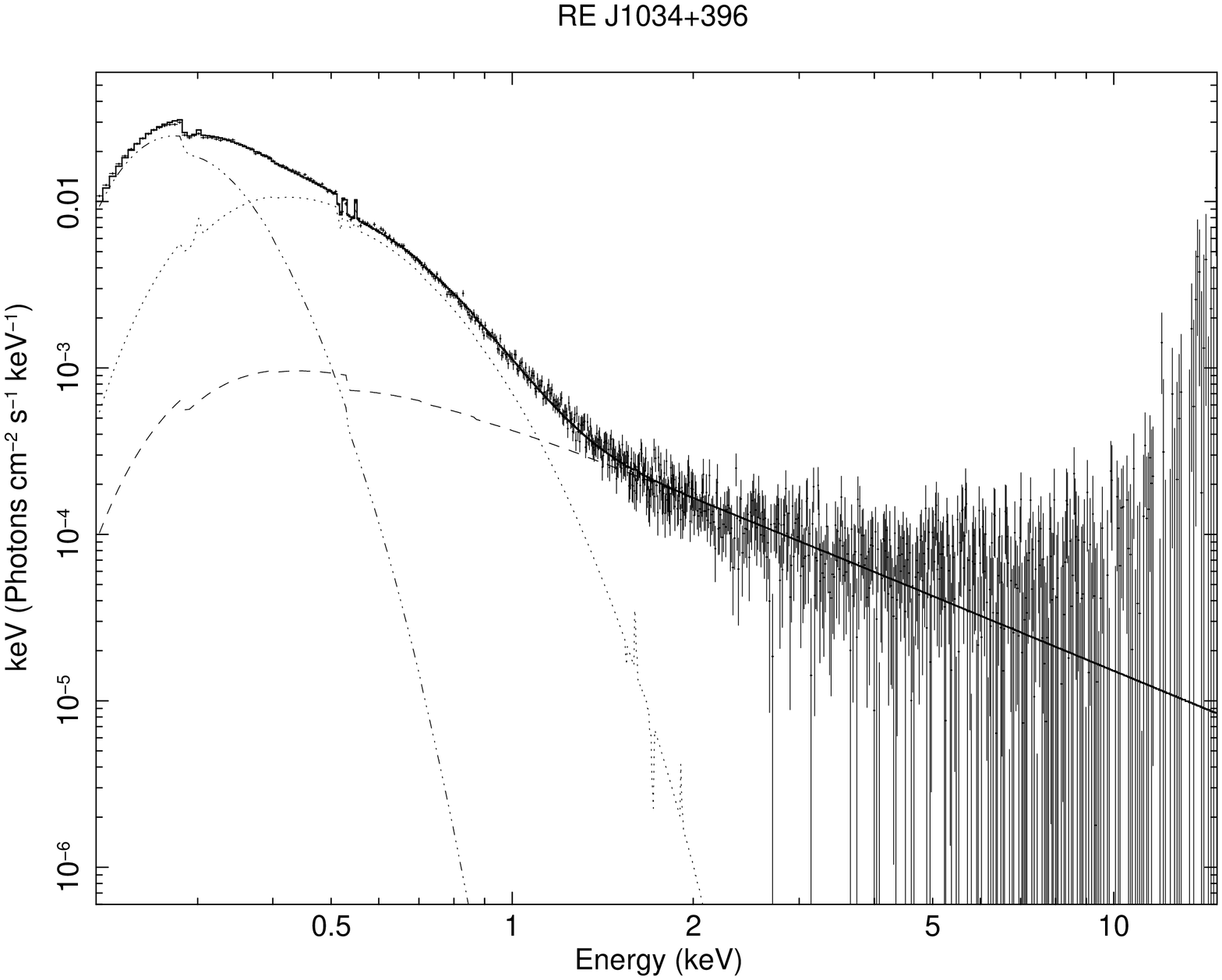}
\includegraphics[scale=0.2,angle=-90]{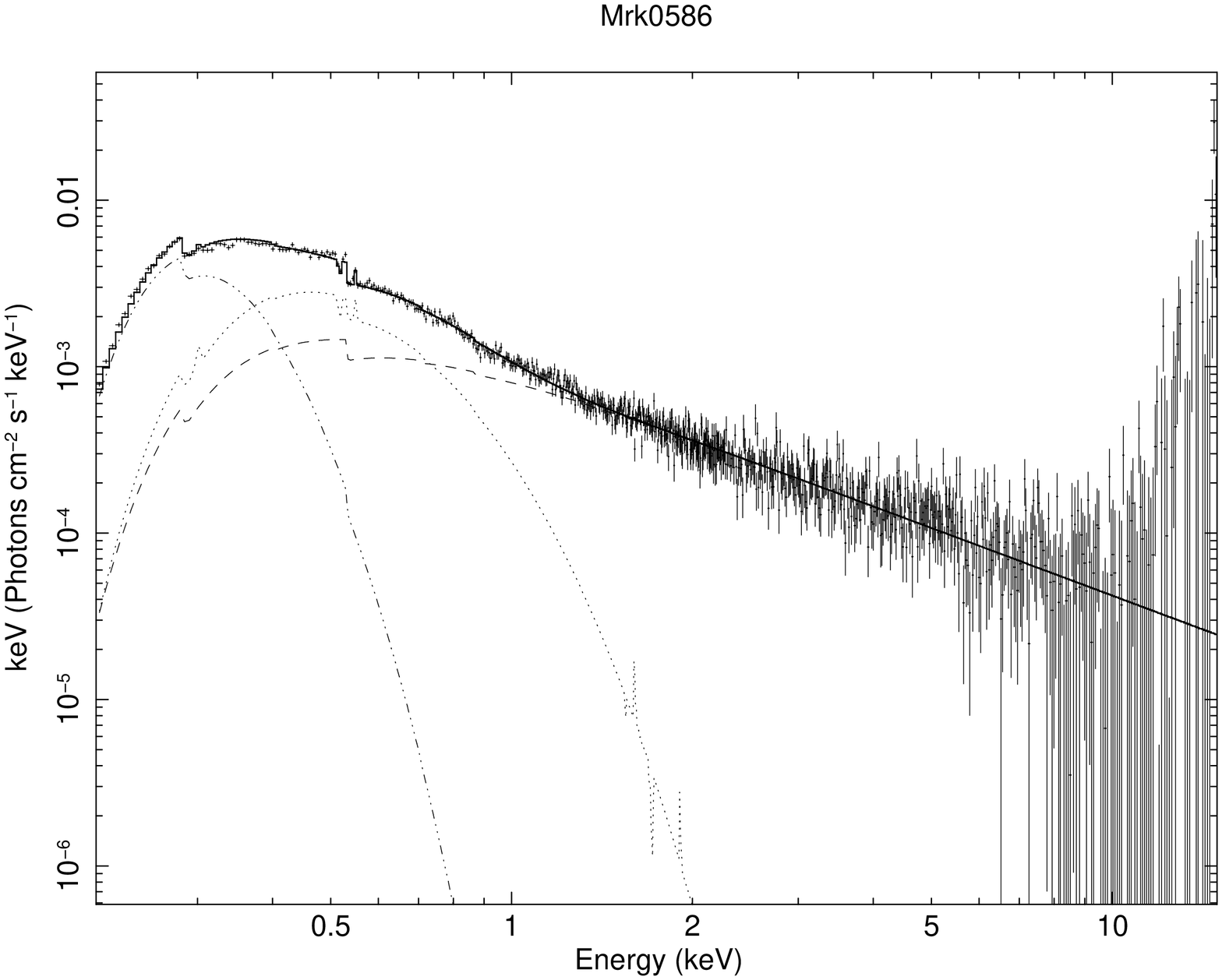}
\caption[spectrum]{\label{Photoz} The fitting spectra with blackbody component. The black crosses are the data points, the solid line is the best fit to data. The dot and dash-dot lines show the inverse Compton scattering component and blackbody component from disc respectively.  The dash line is the power law component from corona.}
\end{figure*}

\end{CJK}

\end{document}